\documentclass[trackchanges,longauthor,twocolumn]{aastex63}

\pdfoutput=1

\hypersetup{urlcolor=blue,citecolor=blue}

\definecolor{darkgreen}{rgb}{0,0.5,0}
\definecolor{purple}{rgb}{1,0,1}

\submitjournal{ApJ}
\usepackage{amsmath}

\shorttitle{Probing the High-energy $\gamma$-ray Pulsar Emission Mechanism}
\shortauthors{Barnard et al.}

\begin{document}

\title{Probing the High-energy $\gamma$-ray Emission Mechanism in the Vela Pulsar via Phase-resolved Spectral and Energy-dependent Light Curve Modeling}

\correspondingauthor{Monica Barnard}
\email{monicabarnard77@gmail.com}

\author[0000-0003-1720-7959]{Monica Barnard}
\affiliation{Centre for Space Research, North-West University, Potchefstroom 2520, South Africa}
\affiliation{Centre for Astro-Particle Physics, University of Johannesburg, Auckland Park, 2092, South Africa}

\author[0000-0002-2666-4812]{Christo Venter}
\affiliation{Centre for Space Research, North-West University, Potchefstroom 2520, South Africa}

\author[0000-0001-6119-859X]{Alice K. Harding}
\affiliation{Theoretical Division, Los Alamos National Laboratory, Los Alamos, NM 58545, USA}

\author[0000-0003-1080-5286]{Constantinos Kalapotharakos}
\affiliation{Astrophysics Science Division, NASA Goddard Space Flight Center, Greenbelt, MD 20771, USA}
\affiliation{Universities Space Research Association (USRA),Columbia, MD 21046, USA}
\affiliation{University of Maryland, College Park (UMDCP/CRESST),College Park, MD 20742, USA}

\author{Tyrel J. Johnson} 
\affiliation{College of Science, George Mason University, Fairfax, VA 22030,\\ resident at Naval Research Laboratory, Washington, DC20375, USA}

\begin{abstract} 
Recent kinetic simulations sparked a debate regarding the emission mechanism responsible for pulsed GeV $\gamma$-ray emission from pulsars. Some models invoke curvature radiation, while other models assume synchrotron radiation in the current-sheet. We interpret the curved spectrum of the Vela pulsar as seen by H.E.S.S.\ II (up to $\sim$100 GeV) and the \emph{Fermi} Large Area Telescope (LAT) to be the result of curvature radiation due to primary particles in the pulsar magnetosphere and current sheet. We present phase-resolved spectra and energy-dependent light curves using an extended slot gap and current sheet model, invoking a step function for the accelerating electric field as motivated by kinetic simulations. We include a refined calculation of the curvature radius of particle trajectories in the lab frame, impacting the particle transport, predicted light curves, and spectra. Our model reproduces the decrease of the flux of the first peak relative to the second one, evolution of the bridge emission, near-constant phase positions of peaks, and narrowing of pulses with increasing energy. We can explain the first of these trends because we find that the curvature radii of the particle trajectories in regions where the second $\gamma$-ray light curve peak originates are systematically larger than those associated with the first peak, implying that the spectral cutoff of the second peak is correspondingly larger. However, an unknown azimuthal dependence of the $E$-field as well as uncertainty in the precise spatial origin of the GeV emission, precludes a simplistic discrimination of emission mechanisms. 
\end{abstract}

\section{Introduction} \label{sec:intro}
The field of pulsar science has been revolutionized by the launch of the \emph{Fermi} Large Area Telescope (LAT; \citealt{Atwood2009}), a high-energy (HE) satellite measuring $\gamma$-rays in the range from 20~MeV to over 300~GeV. The \emph{Fermi} LAT has detected more than 250\footnote{https://confluence.slac.stanford.edu/display/GLAMCOG/\\ Public+List+of+LAT-Detected+Gamma-Ray+Pulsars} $\gamma$-ray pulsars and measured their light curves and spectral characteristics in unprecedented detail \citep{Abdo2010PC1,Abdo2013}. The vast majority of the \emph{Fermi}-detected pulsars display exponentially cutoff power-law spectra with spectral cutoffs around a few GeV.

In the very-high-energy (VHE) band, MAGIC detected pulsations from the Crab pulsar at energies up to 1~TeV \citep{Ansoldi2016}, and H.E.S.S.~II detected pulsed emission from the Vela pulsar in the sub-20 to 100~GeV range \citep{Abdalla2018}. New observations by H.E.S.S.\ reveal pulsed emission from Vela at a few TeV (H.E.S.S.\ Collaboration, in preparation). H.E.S.S.~II furthermore detected pulsed emission from PSR~B1706$-$44 in the sub-100~GeV energy range \citep{SpirJacob2019}. Pulsed emission from the Geminga pulsar between 15 and 75~GeV at a significance of 6.3$\sigma$ was recently detected by MAGIC, although only the second light curve peak is visible at these energies. The MAGIC spectrum (which can be represented by a simple power law) is an extension of the \emph{Fermi} LAT spectrum, ruling out the possibility of a subexponential cutoff in the same energy range at the $3.6\sigma$ level and possibly indicating a transition from a curvature radiation (CR) to an inverse Compton (IC) spectral component \citep{Acciari2020}.
Interestingly, as the photon energy $E_{\gamma}$ is increased (above several GeV), the main light curve peaks of Crab, Vela, and Geminga seem to remain at the same phase positions, the intensity ratio of the first to second peak (P1/P2) decreases for Vela and Geminga, the interpeak ``bridge'' emission evolves for Vela, and the peak widths decrease for Crab \citep{Aliu2011}, Vela \citep{Abdo2010Vela} and Geminga \citep{Abdo2010Geminga}. The P1/P2 vs.\ $E_{\gamma}$ effect was also seen by \emph{Fermi} for a number of pulsars \citep{Abdo2010PC1,Abdo2013}.

In general, multiwavelength pulsar light curves exhibit an intricate structure that evolves with $E_{\gamma}$ \citep[e.g.,][]{Buehler2014}, reflecting the various underlying emitting particle populations and spectral radiation components that contribute to this emission, as well as the local magnetic ($\boldsymbol{B}$) field geometry (as encoded by the local curvature radius) and electric ($\boldsymbol{E}$) field spatial distribution. In addition, Special Relativistic effects modify the emission beam, given the fact that the corotation speeds may reach close to the speed of light $c$ in the outer magnetosphere.

Some traditional physical emission models invoke CR from extended regions within the magnetosphere to explain the HE spectra and light curves. These include the slot gap (SG; \citealp{Arons1983} \citealt{Harding2003}) and outer gap (OG; \citealt{Romani1995,Cheng1986I}) models. However, they fall short of fully addressing global magnetospheric characteristics, e.g., the particle acceleration and pair production, current closure, and radiation of a complex multiwavelength spectrum. Geometric light curve modeling \citep{Dyks2004,Venter2009,Watters2009,Johnson2014,Pierbattista2015} presented an important interim avenue for probing the pulsar magnetosphere in the context of traditional pulsar models, focusing on the spatial rather than physical origin of HE photons. More recent developments include global magnetospheric models such as the force-free (FF; \citealt{Kalapotharakos2009}) inside and dissipative outside (FIDO) model \citep{Brambilla2015,Kalapotharakos2014,Kalapotharakos2017}, equatorial current-sheet models (e.g., \citealt{Bai2010,Contopoulos2010,Petri2012}), the striped-wind models (e.g., \citealt{Petri2011}), and kinetic/particle-in-cell simulations (PIC; \citealt{Brambilla2018,Cerutti2016current,Cerutti2016,Cerutti2020,Kalapotharakos2018,Philippov2018}). Some studies using the FIDO models assume that particles are accelerated by induced $E$-fields in dissipative magnetospheres and produce GeV emission via CR (e.g., \citealt{Kalapotharakos2014,Brambilla2015}). Conversely, in some of the wind or current-sheet models, HE emission originates beyond the light cylinder via synchrotron radiation (SR) by relativistic, hot particles that have been accelerated via magnetic reconnection inside the current sheet \citep[e.g.,][]{Petri2011,Philippov2018}.
Given the ongoing debate between the emission mechanisms of HE emission, our motivation in this paper is to explain the GeV spectrum and light curves of Vela as measured by \emph{Fermi} and H.E.S.S.\ Specifically, by modeling the $E_\gamma$-dependent light curves (and P1/P2 signature) in the CR regime of synchrocurvature (SC) radiation, we hope to probe whether this effect can serve as a potential discriminator between emission mechanisms and models (see also the reviews of~\citealt{Harding2016,Venter2016,Venter2017} on using pulsar light curves to scrutinize magnetospheric structure and emission distribution).

The structure of the rest of the paper is as follows. Focusing on the GeV band, in Section~\ref{sec:model} we briefly discuss a steady-state emission model of \citet{Harding2015} that predicts $E_{\gamma}$-dependent light curves and spectra that result from primary particles emitting CR. We also describe (see Appendix~\ref{sec:AppA}) a refined calculation of the curvature radius $\rho_{\rm c}$ of the particle trajectory, as well as our ``reverse mapping'' method used to isolate the spatial origin of the light curve peaks (\ref{sec:revmap}). In Section~\ref{sec:results} we present sample light curves and spectra, showing the behavior of the peaks as a function of $\rho_{\rm c}$, as applied to the Vela pulsar. For the optimal light curve and spectral fits, we study the local environment of the peaks' emission regions, finding a systematic difference in $\rho_{\rm c}$, particle Lorentz factor $\gamma$, and spectral cutoff energy $E_{\gamma,\rm CR}$ for the two peaks. Our concluding remarks follow in Section~\ref{sec:concl} (these quantities being larger for the second peak). The results of this paper accompany those of \citet{Harding2018}.

\section{Model Description and Improvements} \label{sec:model}
\subsection{A 3D Pulsar Emission Model} \label{subsec:3dmodel}
Using the emission model of \citet{Harding2015} and \citet{Harding2018}, we study the full particle transport but focus on the CR emission component by primary particles in this paper. This model assumes a 3D FF $B$-field as the basic magnetospheric structure. This solution (formally assuming an infinite plasma conductivity, so that the $E$-field is fully screened) serves as a good approximation to the geometry of field lines implied by the dissipative models that require a high conductivity in order to match observed $\gamma$-ray light curves \citep{Kalapotharakos2012,Kalapotharakos2014}.

Both primary particles (leptons) and electron$-$positron pairs are injected at the stellar surface. The primaries radiate CR and some of these $\gamma$-ray photons are converted into pairs in the intense $B$-fields close to the star. The primaries are injected with a low initial speed and are further accelerated along the $B$-field lines by a constant parallel $E$-field $E_{\parallel}$ (used as a free parameter in this model), in an extended SG scenario near the last open field lines. Using an independent code, we calculate a polar cap (PC) pair cascade that develops just above the neutron star surface, since the pairs radiate SR and the primaries CR, leading to further generations of particles with lower energies. In the global model, the resulting pair spectrum is injected at the stellar surface and is responsible for, e.g., a pair SR emission component that may be dominant in the X-ray band, although we do not focus on this particular model output in this paper.

In this model, the SG reaches beyond the light cylinder radius $R_{\rm LC}=c/\Omega$ (where the corotation speed equals $c$, with $\Omega$ the angular speed) up to altitudes of $r=2R_{\rm LC}$. For more details, see \citet{Harding2015}.

\begin{figure*}[t]
\epsscale{0.9}
\plotone{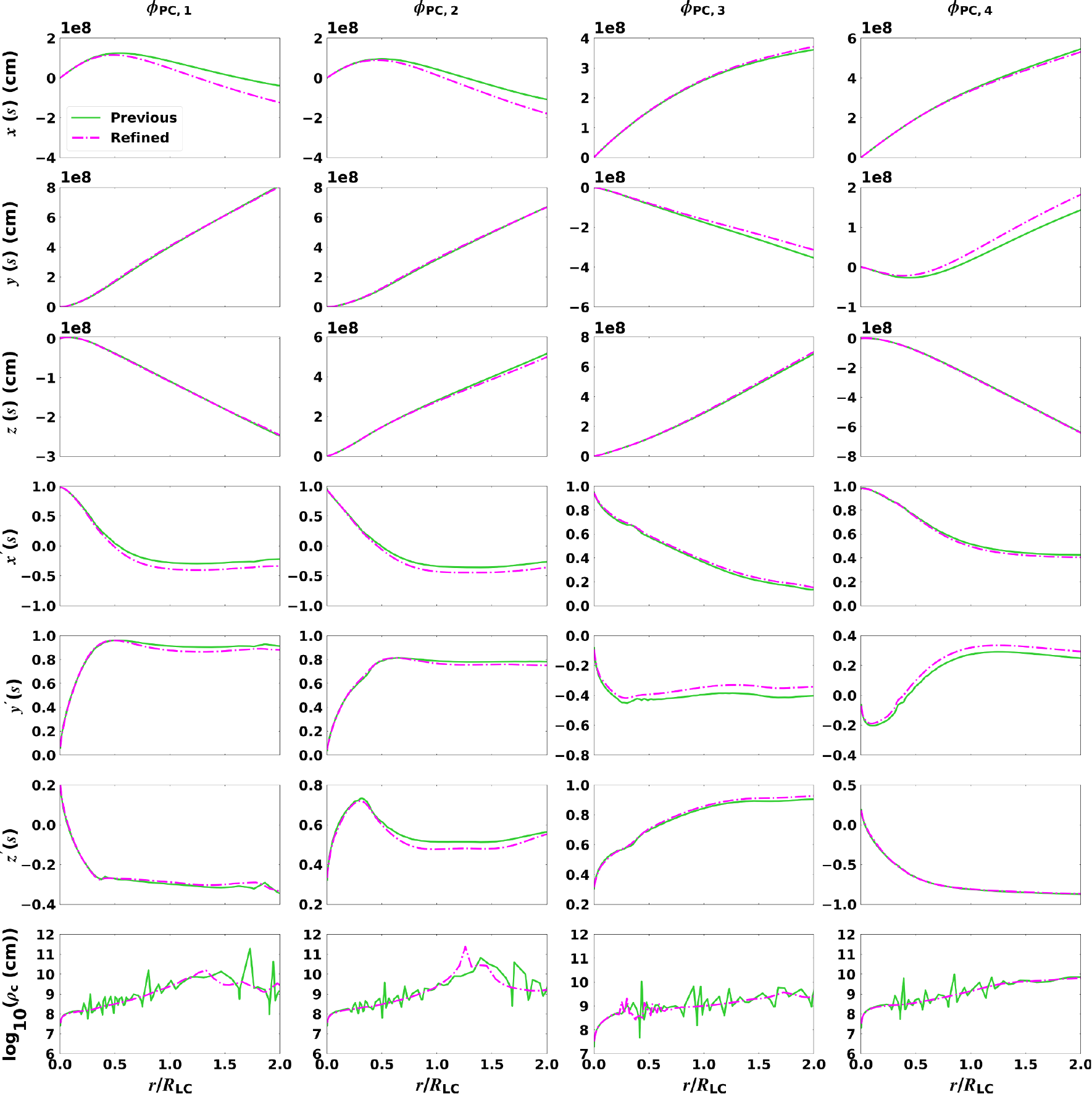}
\caption{A comparison of electron position $x(s)$,\,$y(s)$,\,$z(s)$; trajectory direction $x'(s)$,\,$y'(s)$,\,$z'(s)$; and $\log_{10}(\rho_{\rm c})$, as calculated previously (lime green) and now being refined (magenta), for a magnetic inclination angle $\alpha=75^\circ$, along four arbitrary $B$-field lines (i.e., field-line footpoints with $\phi_{\rm PC,1}=45^\circ$, $\phi_{\rm PC,2}=135^\circ$, $\phi_{\rm PC,3}=225^\circ$ and $\phi_{\rm PC,4}=315^\circ$) on the outer ring of the PC. \label{fig:posdirrhoOvsN}}
\end{figure*}

\subsection{Improved Particle Trajectory Calculations}
\label{subsec:rhocalc}

We refine the previous first-order calculation of $\rho_{\rm c}$ along the electron (or positron) trajectory in the lab frame, assuming that all particles injected at the footpoint of a particular $B$-field line follow the same trajectory, independent of their energy because they are quickly accelerated to relativistic energies by the unscreened $E$-field. This independence of $\rho_{\rm c}$ on energy also reduces computational time significantly, because the calculation is done beforehand. We furthermore assume that the $B$-field is strong enough to constrain the movement of the electrons so they will move parallel to the field line in the corotating frame. Thus, there will be no perpendicular motion in the corotating frame because the perpendicular particle energy is nearly instantly expended via SR. We thus take into account the perpendicular $\boldsymbol{E}\times\boldsymbol{B}$ drift in the lab frame.

To calculate the electron's trajectory as well as its associated $\rho_{\rm c}$ in the lab frame, we used a small, fixed step size $ds$ (where $s$ is the arclength) along the $B$-field line. The first derivative along the trajectory indicates the direction of the particle's longitudinal motion. Next, we smooth the directions using $s$ as the independent variable to counteract numerical noise. 
Second, we match the unsmoothed and smoothed directions of the electron trajectory at particular $s$ values to get rid of unwanted ``tails'' at low and high altitudes, introduced by the use of a Gaussian kernel density estimator (KDE; \citealt{Parzen1962}) smoothing procedure. 
Third, we use a second-order method involving interpolation by a Lagrange polynomial to obtain the second-order derivatives of the positions along the trajectory as a function of $s$ \citep{Faires2002}. This accuracy is necessary because $\rho_{\rm c}$ is a function of second-order derivatives of the electron position, and instabilities may be exacerbated if not dealt with carefully. Lastly, we match the $\rho_{\rm c}$ calculated using smoothed and unsmoothed directions to get rid of ``tails'' in $\rho_{\rm c}$ at low and high altitudes, as before. See Appendix~\ref{sec:AppA} for a more detailed discussion and calculations. Having a precalculated $\rho_{\rm c}$ in hand, for a fine division in $s$ along any particular $B$-field line, we then interpolate $\rho_{\rm c}$ in the particle transport calculations to accommodate an adaptive, variable-$ds$ approach that is used to speed up the transport calculations, without losing accuracy of the trajectory. 

In Figure~\ref{fig:posdirrhoOvsN} the parameters describing the particle trajectory are compared for the previous and the newly calculated $\rho_{\rm c}$. These include the particle positions $x(s)$, $y(s)$, $z(s)$ in centimeter; dimensionless directions or spatial derivatives with respect to arclength $x'(s)$, $y'(s)$, $z'(s)$; and the $\log_{10}$ of $\rho_{\rm c}$. This comparison is shown for four arbitrary $B$-field lines with footpoints along the outer ring (rim) on the PC, as indicated by different values of the magnetic azimuth $\phi_{\rm PC}$. The changes in position and direction are rather minor. However, the improved calculation smooths out some instabilities in $\rho_{\rm c}(s)$. 

\begin{figure}
\epsscale{1.2}
\plotone{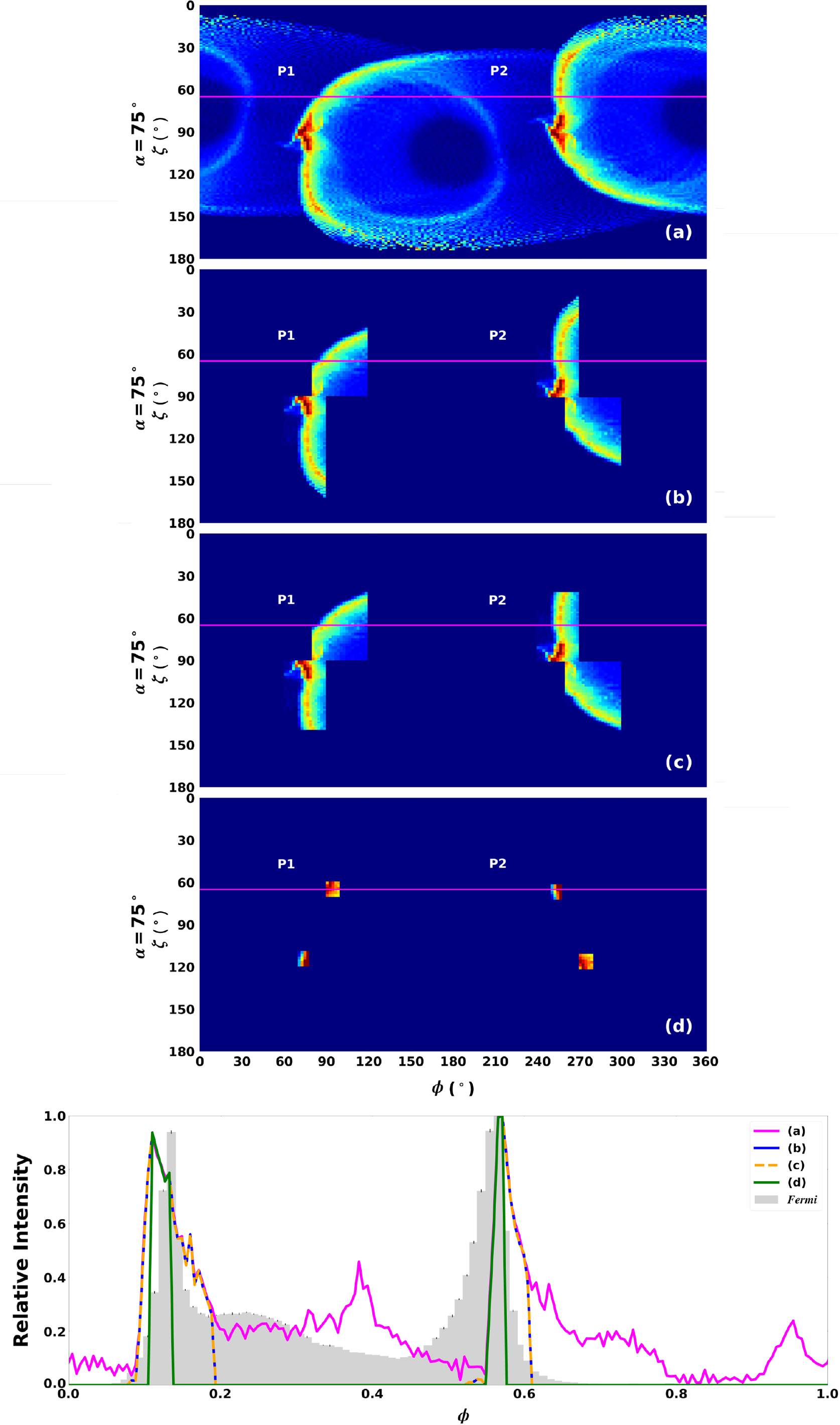}
\caption{Example phase plots with $(\phi,\zeta)$ ``blocks'' (or 2D bins) and their associated light curves for $\alpha=75^\circ$, $\zeta_{\rm cut}=65^\circ$, $R_{\rm acc}=eE_{\parallel}/{m_{\rm e}}c^2=0.25$~cm$^{-1}$, and $0.1<E_\gamma<50$~GeV. In order to indicate how we isolated the first and second light curve peaks (labeled `P1' and `P2'), we made cuts in $(\phi,\zeta)$ as follows: (a) no cut, (b) $\phi_{\rm P1}=[-100^\circ,-60^\circ]$ and $\phi_{\rm P2}=[-120^\circ,-90^\circ]$ for all $\zeta$,
(c) $\zeta_{\rm P1}=[90^\circ,160^\circ]$ and $\zeta_{\rm P2}=[40^\circ,90^\circ]$ for same $\phi$ as in (b), and 
(d) $\phi_{\rm P1}=[-90^\circ,-81^\circ]$, $\zeta_{\rm P1}=[110^\circ,120^\circ]$, $\phi_{\rm P2}=[-109^\circ,-103^\circ]$, and $\zeta_{\rm P2}=[60^\circ,70^\circ]$. The light curve legend in the lower panel refers to each associated phase plot, for increasingly smaller $(\phi,\zeta)$ bins. The \textit{Fermi} data for Vela are indicated by a gray histogram (\citealt{Abdo2013},\url{http://fermi.gsfc.nasa.gov/ssc/data/access/lat/2nd_PSR_catalog/}). We shifted the resulting $\gamma$-ray model light curves by $-0.14$ in normalized phase to fit the data. This reflects the degeneracy of $\phi=0$ in the data (reflecting the main radio peak) and $\phi=0$ (the phase of the magnetic axis). \label{fig:PPLCblocks}}
\end{figure}

\section{Isolating the Spatial Origin of Emission for Each of the Light Curve Peaks} \label{sec:revmap}

As mentioned in Section~\ref{sec:intro}, the relative fading of peak~1 vs.\ peak~2 with $E_\gamma$ seems to be a common characteristic of HE light curves. We have also been able to reproduce this with the code. In order to probe the origin of this effect, it is necessary to isolate the spatial origin of each light curve peak and study key parameters in these regions. We start by isolating each peak on the phase plot (see definition below; using increasingly smaller observer angle $\zeta$ and rotation phase $\phi$ bins) and then apply ``reverse mapping'' to uncover the emission's spatial position. This can be compared to developing a ``reverse dictionary'' that translates a chosen $(\phi,\zeta)$ range into a spatial range within the magnetosphere.

Using the model described in Section~\ref{subsec:3dmodel} and
for a given magnetic inclination angle $\alpha=75^\circ$, we generated phase plots (see Figure~\ref{fig:PPLCblocks}). We perform simulations for the Vela pulsar for the following parameters.\footnote{\citet{Manchester2005}: spin period $P=0.089$~ms, its time derivative $\dot{P}=1.25\times10^{-13}$~s$^{-1}$, and $d=0.29$~kpc.}. We inject the primaries into a roughly annular SG situated between $r_{\rm ovc}=0.90$ and $r_{\rm ovc}=0.96$ (in units of PC radius; \citealt{Dyks2004,Harding2018}), and divide the cross section of the surface projection of the SG situated near the rim of the PC into 7 rings, with each ring having $360$ azimuthal segments. We additionally set $ds=10^{-3}R_{\rm LC}$ with a corresponding KDE smoothing parameter $h=50ds$. The phase plots are emitted photon fluxes $\dot{N}_{\gamma}$ that have been normalized using the primary particle flux (the appropriate Goldreich$-$Julian injection rate at the stellar surface); $\dot{N}_{\gamma}$ is collected in bins of $\zeta$ and $\phi$. The photon phases have been corrected for rotation and time-of-flight delays. Lastly, $\dot{N}_{\gamma}$ per bin is divided by the solid angle subtended by each phase plot bin, i.e., $\delta\Omega=(\cos\zeta-\cos(\zeta+\delta\zeta))\delta\phi\approx\sin\zeta{\delta\zeta}{\delta\phi}$, and the energy bin width $dE_\gamma$. To generate light curves, a constant-$\zeta$ cut ($\zeta_{\rm cut}$) is made through the respective phase plot (see the lower panel of Figure~\ref{fig:PPLCblocks}), integrating over some photon energy range. On the other hand, to generate spectra, we also make a constant-$\zeta$ cut; however, for each fixed photon energy, we now integrate over $\phi$ (see Appendix~\ref{sec:AppA}).

In our code, we calculate the emission from the northern rotational hemisphere $\dot{N}^\prime_\gamma(\phi,\zeta)$ only. The contribution of the emission from the southern hemisphere $\dot{S}_\gamma^\prime$ is obtained by taking into account the symmetry with respect to the center of the star (i.e., $\dot{S}^\prime_{\gamma} =\dot{N}^\prime_{\gamma}(\phi+180^{\circ},180^{\circ}-\zeta)$). This symmetry exploitation saved computational time and the total emission is then given by $\dot{N}^\prime_\gamma+\dot{S}^\prime_\gamma$. The corresponding full sky map (from which the spectra and light curves are derived) is shown in the first panel of Figure~\ref{fig:PPLCblocks}. The implication for calculations of, e.g., the histogram of the local values of $\rho_{\rm c}$ as done in Section~\ref{sec:results}, is that one has to carefully keep track of the ($\phi,\zeta$) coordinates of each peak and map them back onto the northern-hemisphere caustic (e.g., mapping P1 onto $\dot{N}^\prime_\gamma$ where $\zeta>90^\circ$). Using the latter prescription, one can perform the reverse mapping to find the spatial coordinates of the emission associated with each peak. Thus, our spectra and light curves are based on two, overlapping caustics; however, we make an approximation when studying the local $\rho_{\rm c}$ distribution associated with each peak by only taking $\dot{N}^\prime_\gamma$ into account because this will far outweigh any low-level contribution from $\dot{S}^\prime_\gamma$.

This reverse mapping procedure is illustrated in Figure~\ref{fig:PPLCblocks} for a constant acceleration ``rate'' (acceleration per unit length) $R_{\rm acc}=eE_{\parallel}/{m_{\rm e}}c^2$~cm$^{-1}$, with $e$ the electron charge, $m_{\rm e}$ the electron mass, and $m_{\rm e}c^2$ the rest-mass energy. The first panel is for the full phase space, whereas panels (b), (c), and (d) are for different $(\phi,\zeta)$ ``blocks'' or bins. In panel (b), we make a cut in $\phi$ for both peaks but keep $\zeta$ fixed and see that only the peaks remain on the corresponding light curve (see bottom panel). If we then narrow the range in $\zeta$ for a fixed $\phi$ interval (same as in (b)), we note that the light curve remains the same. Lastly, we make $\phi$ and $\zeta$ small enough so that only the maximum of each peak is included in the $(\phi,\zeta)$ range, as seen in panel (d). These ranges in $\phi$ and $\zeta$ are referred to as the ``optimal bins'' for both peaks and are necessary for constructing the phase-resolved spectra of each peak. We chose the $\zeta$-range for each peak with width of $\pm{5^\circ}$ around $\zeta_{\rm cut}=65^\circ$, to include the $\zeta$ inferred from the pulsar wind nebula torus fit of Vela \citep{Ng2008}.

\begin{figure}
\centering
\epsscale{1.2}
\plotone{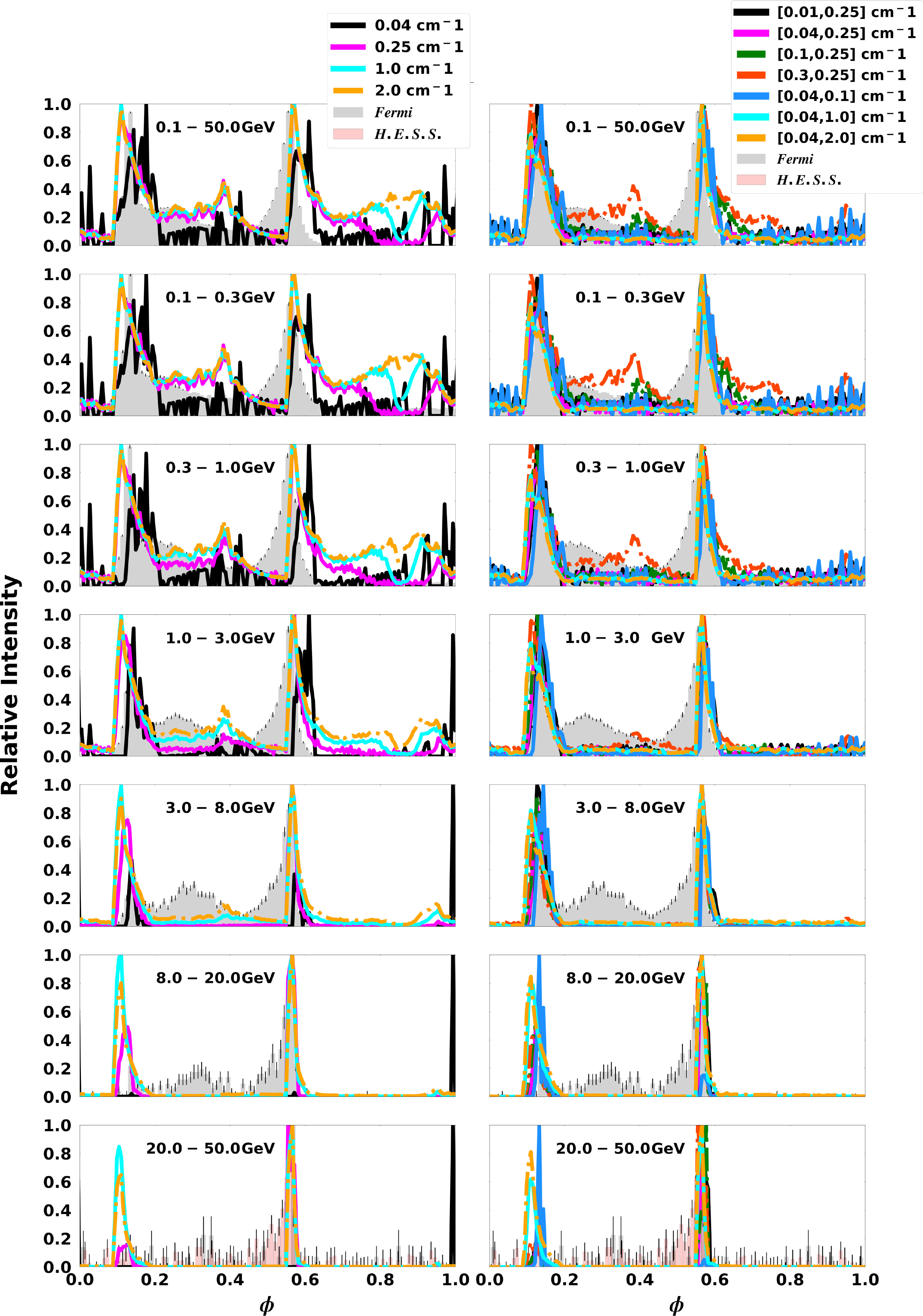}
\caption{Energy-dependent light curves for $\alpha=75^\circ$ and $\zeta_{\rm cut}=65^\circ$ for several different combinations of $R_{\rm acc}$ for both the constant $E_\parallel$ (left column) and two-valued $E_\parallel$ (right column) case. The top panels are for the full $E_\gamma$-range, and for each panel thereafter, the minimum $E_\gamma$ is increased as indicated. We are fitting the model light curves to the \emph{Fermi} (\citealt{Abdo2010Vela,Abdo2013},\url{http://fermi.gsfc.nasa.gov/ssc/data/access/lat/2nd_PSR_catalog/}), and H.E.S.S.\ (at $E_\gamma>20$~GeV; \citealt{Abdalla2018}) data points. We shifted the predicted light curves by $\delta=-0.14$ in normalized phase. One observes that for some choices of $E_\parallel$, the P1/P2 decrease with $E_\gamma$ is more apparent than for others. \label{fig:LC_epardep}}
\end{figure}

\begin{figure}
\centering
\epsscale{1.2}
\plotone{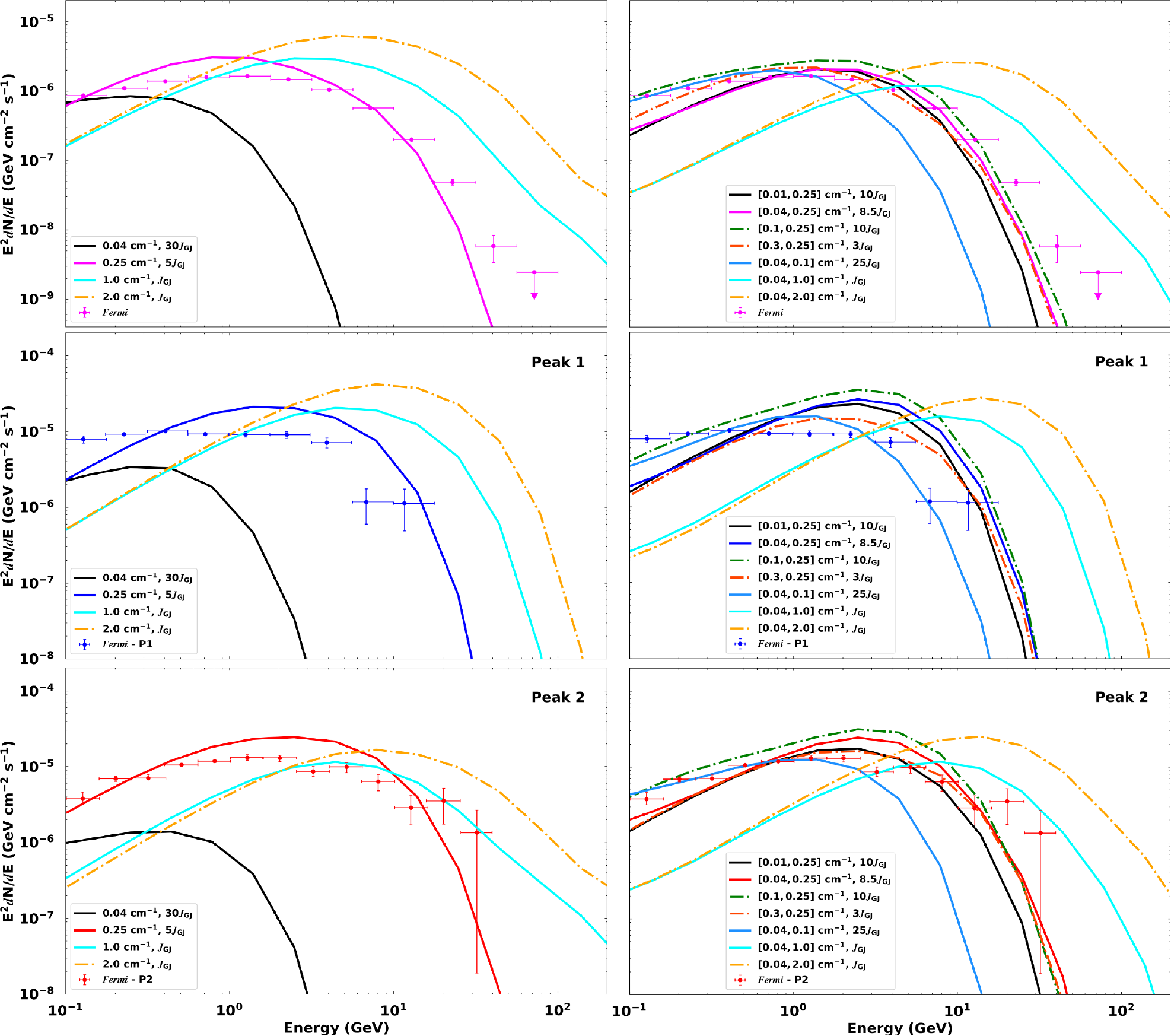}
\caption{Model phase-averaged (top panels) and phase-resolved (middle and bottom panels) spectra associated with Figure~\ref{fig:LC_epardep} for the same for $\alpha$, $\zeta_{\rm cut}$ and $R_{\rm acc}$-field combinations, for both scenario 1 (left column) and scenario 2 (right column). In each $E_\parallel$ case the legend indicates the chosen values for $R_{\rm acc}$, $R_{\rm acc,\rm low}$, $R_{\rm acc,\rm high}$, and the flux normalization factor. The data points for the phase-average spectra are from \citet{Abdo2013} (see \url{http://fermi.gsfc.nasa.gov/ssc/data/access/lat/2nd_PSR_catalog/}) and the phase-resolved spectra are from \citet{Abdo2010Vela}; see text for details.}  \label{fig:spec_epardep}
\end{figure}

\section{Results} \label{sec:results}
\subsection{Finding Optimal Fitting Parameters} \label{subsec:optparams}
After having isolated the spatial origin of the emission of each light curve peak as described in the previous section, we first perform joint light curve and spectral fitting to find optimal model parameters; subsequently, we will consider the local environments where the respective light curve peaks originate (Section~\ref{sec:results}), given these optimal parameters.

We consider two cases throughout this paper, based on either a constant or a two-step parametric accelerating $E_\parallel$-field, independent of the $\phi_{\rm PC}$, $\zeta$ and $r$. Thus, we choose (and subsequently refer to this as scenario~1 and scenario~2): (1) a constant $R_{\rm acc}$ from the stellar surface and into the current sheet (see \citealt{Harding2015}), and (2) a two-valued $R_{\rm acc}$, where $R_{\rm acc,\rm low}$ occurs inside and $R_{\rm acc,\rm high}$ outside the light cylinder (see \citealt{Harding2018}). The two-step function for the accelerating $E_\parallel$ is motivated by global dissipative models \citep{Kalapotharakos2014,Kalapotharakos2017,Brambilla2015} and kinetic PIC models \citep{Cerutti2016,Kalapotharakos2018}.

We performed a preliminary parameter study to search for an optimal combination of $\alpha$, $\zeta_{\rm cut}$, and $R_{\rm acc}$ (for both scenarios, respectively), calibrated against \textit{both} the observed HE light curves \textit{and} spectra measured by \emph{Fermi} and H.E.S.S.~II. We start (for $\alpha=75^\circ$) by fixing $\zeta_{\rm cut}$ and testing different values of $R_{\rm acc}$; later, we fix $R_{\rm acc}$ and free $\zeta_{\rm cut}$\footnote{Given how computationally expensive this exercise is, we only considered a few values of the free parameters. In future, a more robust method may be considered where parameter space of several free parameters may be searched for optimal joint light curve and spectral fits. Given the disparate nature of these data, and the complexity of such a joint fit, here we perform a pilot study to indicate the effect of the different parameters, and to find a reasonable joint fit by eye.}.

Figure~\ref{fig:LC_epardep} shows the $E_\gamma$-dependent light curves for scenario~1 (left column) and scenario~2 (right column). For scenario~1, we choose four arbitrary constant $R_{\rm acc}$ values, and for scenario~2, seven arbitrary $R_{\rm acc, \rm low}$ and $R_{\rm acc, \rm high}$ combinations, as indicated in the legends. We also indicate different energy ranges (with the minimum $E_\gamma$ increasing from top to bottom), with the first panel showing light curves for a full HE range $E_\gamma\in(100~{\rm MeV},50~{\rm GeV})$. The $E_\gamma$ ranges correspond to those of the \emph{Fermi} light curves in Figure~2 in \citet{Abdo2010Vela}, and \citet{Abdo2013}, as well as $E_\gamma>$20~GeV to match the H.E.S.S.~II data \citep{Abdalla2018}. 
The light curves for scenario~1 display bridge emission at $\phi\geq0.25$ that diminishes as $E_\gamma$ increases. For scenario~2, bridge emission develops when $R_{\rm acc, \rm low}\geq0.10$~cm$^{-1}$ and $R_{\rm acc, \rm high}=0.25$~cm$^{-1}$. For $R_{\rm acc, \rm low}=0.3$~cm$^{-1}$ and $R_{\rm acc, \rm high}=0.25$~cm$^{-1}$ the light curve almost mimics our fit in scenario~1 for $R_{\rm acc}=0.25$~cm$^{-1}$ (because these respective values are so close). If both $R_{\rm acc, \rm low}$ and $R_{\rm acc, \rm high}$ are small, we obtain light curve shapes that are contrary to what is expected, e.g., choosing $R_{\rm acc, \rm low}=0.04$~cm$^{-1}$ and $R_{\rm acc, \rm high}=0.1$~cm$^{-1}$, yields an increase of P1/P2 for at $E_\gamma>8.0$~GeV, contrary to what is observed. This may be linked to the fact that the particles do not reach the CR radiation reaction limit in such a case. The optimal choice in terms of reproducing the P1/P2 effect seems to be $(R_{\rm acc, \rm low}$,$R_{\rm acc, \rm high}) = (0.04,0.25)$~cm$^{-1}$, although the bridge emission is somewhat underpredicted.

In both scenarios, four main trends are evident in our optimal fits to the light curves as they evolve with $E_\gamma$. First, the model peaks remain at the same phase, i.e., P1 at $\phi$=[0.10,0.18] and P2 at $\phi$=[0.57, 0.60], after we shifted the model in phase by $\delta=-0.14$ to fit the data (we focused on the $\gamma$-ray data only and do not take the radio peak position into account in this study). Second, the intensity ratio of P1 relative to P2 decreases as $E_{\gamma}$ increases in some cases, where the peaks are nearly equal in height at lower $E_{\gamma}$. Third, the bridge emission fades at higher energies, possibly reflecting its softer spectrum and its origin at lower altitudes, where acceleration is suppressed as compared to the current-sheet environment. Lastly, the pulse width decreases with an increase in $E_\gamma$. It is encouraging that the model can broadly reproduce these observational trends. We also note that a two-step $E_\parallel$-field provides more reasonable light curve shapes closer to the observed ones, especially at lower photon energies.

The observed phase-averaged CR spectra are characterized by a power law with a (sub)exponential cutoff. In our model, this spectrum is calculated as the observed $\dot{N}_{\gamma}$ at a particular viewing angle $\zeta_{\rm cut}$, summing the fluxes (originating in different parts of the magnetosphere) over $\phi$ and dividing by $2\pi{d}^2$, where $d$ (in centimeters) is the distance to the source (see Appendix~\ref{sec:AppB}). To calculate the phase-resolved spectra associated with each peak, we limit the $\phi$-range to include the specific fraction of the emission we want to study. We scaled the phase-resolved flux with the ratio of the relevant peak's $\phi$-range of the \emph{Fermi} data to that of the model range, ensuring that we have comparable quantities in terms of flux per unit phase bin. Figure~\ref{fig:spec_epardep} shows the phase-averaged and phase-resolved (for both P1 and P2) spectra per row. These are associated with the light curves in Figure~\ref{fig:LC_epardep}, for both scenarios and the same parameter values as in Figure~\ref{fig:LC_epardep}. The phase-resolved spectra are taken from \citet{Abdo2010Vela}. Since the predicted CR $\dot{N}_\gamma$ are lower than the \emph{Fermi} data points (\citealt{Abdo2010Vela,Abdo2013}, \url{http://fermi.gsfc.nasa.gov/ssc/data/access/lat/2nd_PSR_catalog/}), we scaled the model with a flux normalization factor in terms of $J_{\rm GJ}$. The flux normalization factor is a multiple of the Goldreich$-$Julian current density $J_{\rm GJ}=\rho_{\rm GJ}c$ (with $\rho_{\rm GJ}=-(\boldsymbol{\Omega}\cdot\boldsymbol{B})/2\pi c$ the corresponding charge density; \citealt{Goldreich1969}). This spectrum normalization has some freedom because the actual current composition (local multiplicity of the high-energy particles) in the pulsar magnetosphere is not absolutely certain. In the figure legend, we indicate $[R_{\rm acc},J_{\rm GJ}]$ for scenario~1 and $[R_{\rm acc,\rm low},R_{\rm acc,\rm high},J_{\rm GJ}]$ for scenario~2.

For scenario~1, at small $R_{\rm acc}$ the model does not fit the data, given the predicted flux distribution (too low a spectral cutoff) with energy. This may be addressed in the future by invoking SC emission, rather than pure CR \citep{Harding2018}. As $R_{\rm acc}$ increases, the model better fits the data; however, when it becomes too large, it shifts the spectra to larger $E_\gamma$'s and the spectral shape changes and deviates from the data points. This reflects the fact that a larger accelerating $E$-field is implied, leading to a larger particle energy and spectral cutoff. Also, for larger $R_{\rm acc}$ the flux normalization factor becomes smaller. This flux factor should in principle be constant for the phase-averaged and phase-resolved spectra, but the flux level is not consistent between the different predicted spectra, e.g., at $R_{\rm acc}=0.25$, P1's model spectra overestimate the data, but not for P2 or the phase-averaged spectra. This may point to the need for a spatially dependent normalization of the current in the future.
For scenario~2, most combinations of $R_{\rm acc,\rm low}$ and $R_{\rm acc,\rm high}$ yield a good fit to the data, except when both $R_{\rm acc,\rm low}$ and $R_{\rm acc,\rm high}$ are small, e.g., $R_{\rm acc}=[0.04,0.1]$~cm$^{-1}$, or $R_{\rm acc,\rm high}$ is high, e.g., $R_{\rm acc}=[0.04,2.0]$~cm$^{-1}$. When $R_{\rm acc,\rm low}$ is small and $R_{\rm acc,\rm high}$ is very high, the spectra extend to unreasonably high $E_{\gamma}$. For $R_{\rm acc,\rm low}=0.3$~cm$^{-1}$ and $R_{\rm acc,\rm high}=0.25$~cm$^{-1}$ the spectral fits almost mimic our fits in scenario~1 for $R_{\rm acc}=0.25$~cm$^{-1}$, although the flux normalization is a bit lower for scenario~2. In both scenarios, $E_{\gamma,\rm CR}$ varies significantly as we change the parameters, so that for certain choices of $R_{\rm acc}$, P1 may have a larger cutoff than P2, contrary to what is observed. Thus, we settle on $R_{\rm acc}=0.25$~cm$^{-1}$ for scenario~1, and $R_{\rm acc,\rm low}=0.04$~cm$^{-1}$ and $R_{\rm acc,\rm high}=0.25$~cm$^{-1}$ for scenario~2 as optimal values for this paper.

\begin{figure}
\centering
\epsscale{1.1}
\plotone{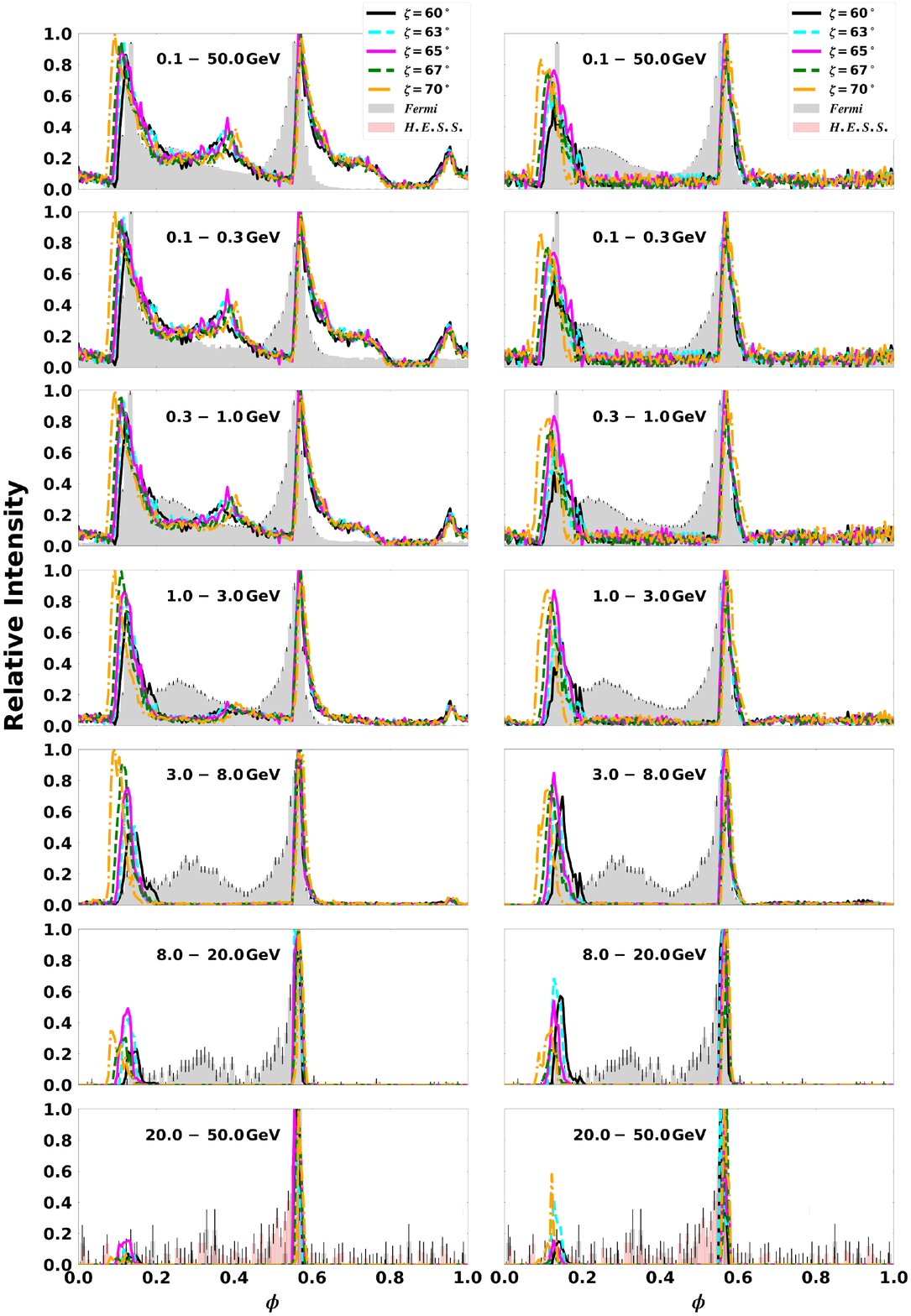}
\caption{Energy-dependent light curves for $\alpha=75^\circ$ and different $\zeta_{\rm cut}$ for the optimal values of the $E_{\parallel}$-field for both scenario~1 (left) and scenario~2 (right). In each $R_{\rm acc}$ case, the legend indicates the chosen values for $\zeta_{\rm cut}$. The first row is for the full $E_\gamma$-range, and each panel thereafter is for an increase in the minimum $E_\gamma$. We are fitting the model light curves to the \emph{Fermi} (\citealt{Abdo2010Vela,Abdo2013},\url{http://fermi.gsfc.nasa.gov/ssc/data/access/lat/2nd_PSR_catalog/}) and to the H.E.S.S.\ (at $E_\gamma>20$~GeV; \citealt{Abdalla2018}) data points, with $\delta=-0.14$. \label{fig:LC_zdep}}
\end{figure}

\begin{figure}
\centering
\epsscale{1.2}
\plotone{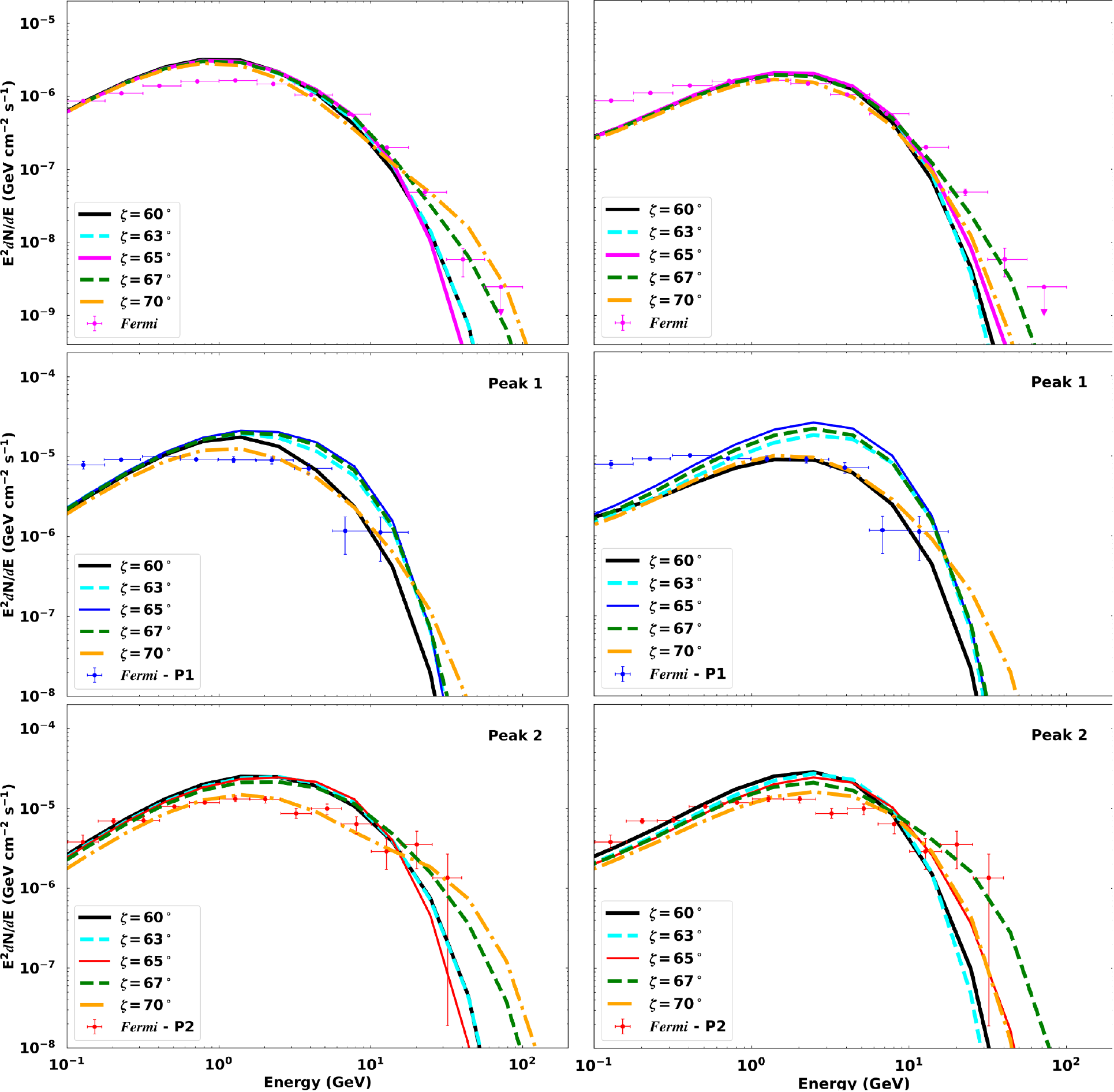}
\caption{Model phase-averaged (top panels) and phase-resolved (middle and bottom panels) spectra associated with Figure~\ref{fig:LC_zdep} for the same $\alpha$, $\zeta_{\rm cut}$ and optimal $R_{\rm acc}$ choices, for both scenario~1 (left) and scenario~2 (right). In each case, the legend indicates the chosen values for $\zeta_{\rm cut}$. The flux normalization factor is $5J_{\rm GJ}$ for the first case and $8.5J_{\rm GJ}$ for the second. The data points for the phase-average spectra are from \citet{Abdo2013} (see \url{http://fermi.gsfc.nasa.gov/ssc/data/access/lat/2nd_PSR_catalog/}), and the phase-resolved spectra are from \citet{Abdo2010Vela}. \label{fig:spec_zdep}}
\end{figure}

Next, we consider the impact of different values of $\zeta_{\rm cut}$ on the predicted light curves and spectra, for the optimal values of $R_{\rm acc}$. Figure~\ref{fig:LC_zdep} indicates energy-dependent light curves for $\alpha=75^\circ$ and $\zeta_{\rm cut}=[60^\circ,63^\circ,65^\circ,67^\circ,70^\circ]$. We notice that P1/P2 decreases with energy at different rates. For larger $\zeta_{\rm cut}$ (i.e., 67$^\circ$ and 70$^\circ$), P1 is relatively higher at lower $E_\gamma$. In scenario~2, the same happens at larger $\zeta_{\rm cut}$ but only at $E_\gamma\geq{20}$~GeV. Also, the level of bridge emission depends on the choice of $\zeta_{\rm cut}$. Figure~\ref{fig:spec_zdep} indicates spectra for the same optimal $R_{\rm acc}$ parameters, but for different $\zeta_{\rm cut}$ values. For smaller $\zeta_{\rm cut}$, the model spectra fit the data well, but for larger $\zeta_{\rm cut}$, the model spectral cutoffs extend to higher $E_\gamma$, sometimes overshooting the data. Also, these spectra are lower in flux than those for the smaller $\zeta_{\rm cut}$ fits (we fixed the flux normalization for all values of $\zeta_{\rm cut}$). 
In scenario~2, the spectral cutoff $E_{\gamma,\rm CR}$ varies significantly, so that for certain choices of $\zeta_{\rm cut}$, P1 has a larger cutoff than P2. If we analyze Figure~\ref{fig:LC_zdep} and~\ref{fig:spec_zdep} concurrently, our optimal fit for both scenarios is for $\zeta_{\rm cut}=65^\circ$. We therefore construct all subsequent figures, e.g., phase plots, light curves, and spectra for optimal values of $\alpha=75^\circ$, and $\zeta_{\rm cut}=65^\circ$ (we indicate spectra for $\alpha=60^\circ$ for comparison). Additional optimal values are $R_{\rm acc}=0.25$~cm$^{-1}$ and a flux normalization factor of $5J_{\rm GJ}$ for scenario~1, and $R_{\rm acc,\rm low}=0.04$~cm$^{-1}$, $R_{\rm acc,\rm high}=0.25$~cm$^{-1}$, and $10J_{\rm GJ}$ for scenario~2. These values produce good fits to the \emph{Fermi} and H.E.S.S.\ II data.




\subsection{Phase Plots, Light Curves and Spectra for the Optimal and Nonoptimal Parameters} \label{subsec:PPLCs}
\begin{figure*}
\epsscale{1.2}
\plotone{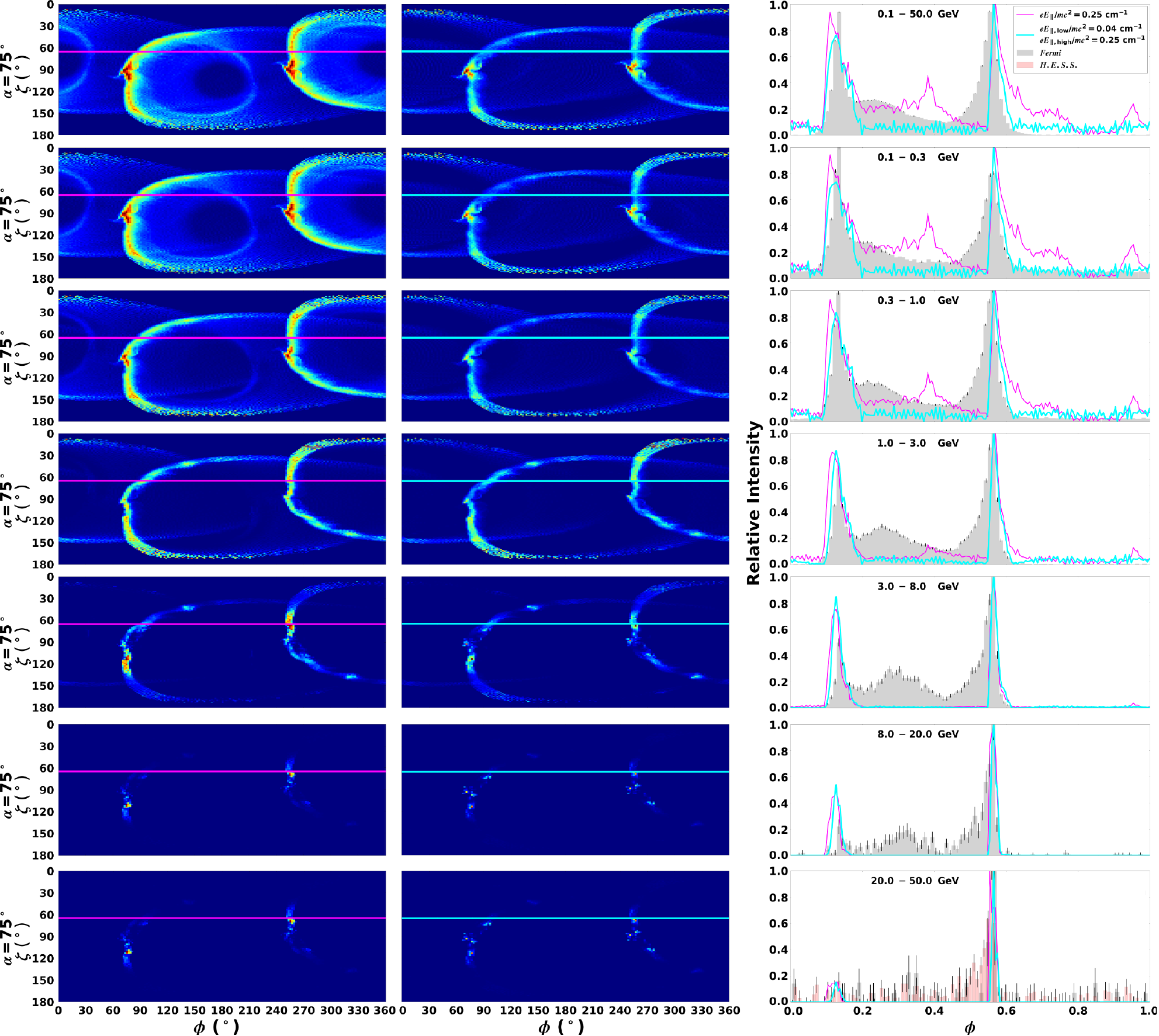}
\caption{Energy-dependent phase plots and light curves for $\alpha=75^\circ$ and $\zeta_{\rm cut}=65^\circ$ and for the optimal $R_{\rm acc}$ for both the first (left column) and second (center column) scenarios, plus their associated light curves (right column). The top panels are for the full $E_\gamma$-range, and each panel thereafter is for a different subband, as indicated by the labels in the light curve panels. Peaks were shifted by $-0.14$ to fit the \textit{Fermi} LAT and H.E.S.S.\ data. \label{fig:PPLCopt}}
\end{figure*}

In Figure~\ref{fig:PPLCopt} we show the energy-dependent phase plots and accompanying light curves for our optimal fit for both scenarios. For scenario~1 (left phase plot) the bridge and most of the off-peak emission disappear with increasing $E_\gamma$, although the light curve peak positions for both scenarios remain roughly stable. The other light curve trends mentioned in Section~\ref{subsec:optparams} are also visible here, i.e., the decrease of P1/P2 and a decrease in peak width with $E_\gamma$.

\begin{figure*}
\epsscale{1.2}
\plotone{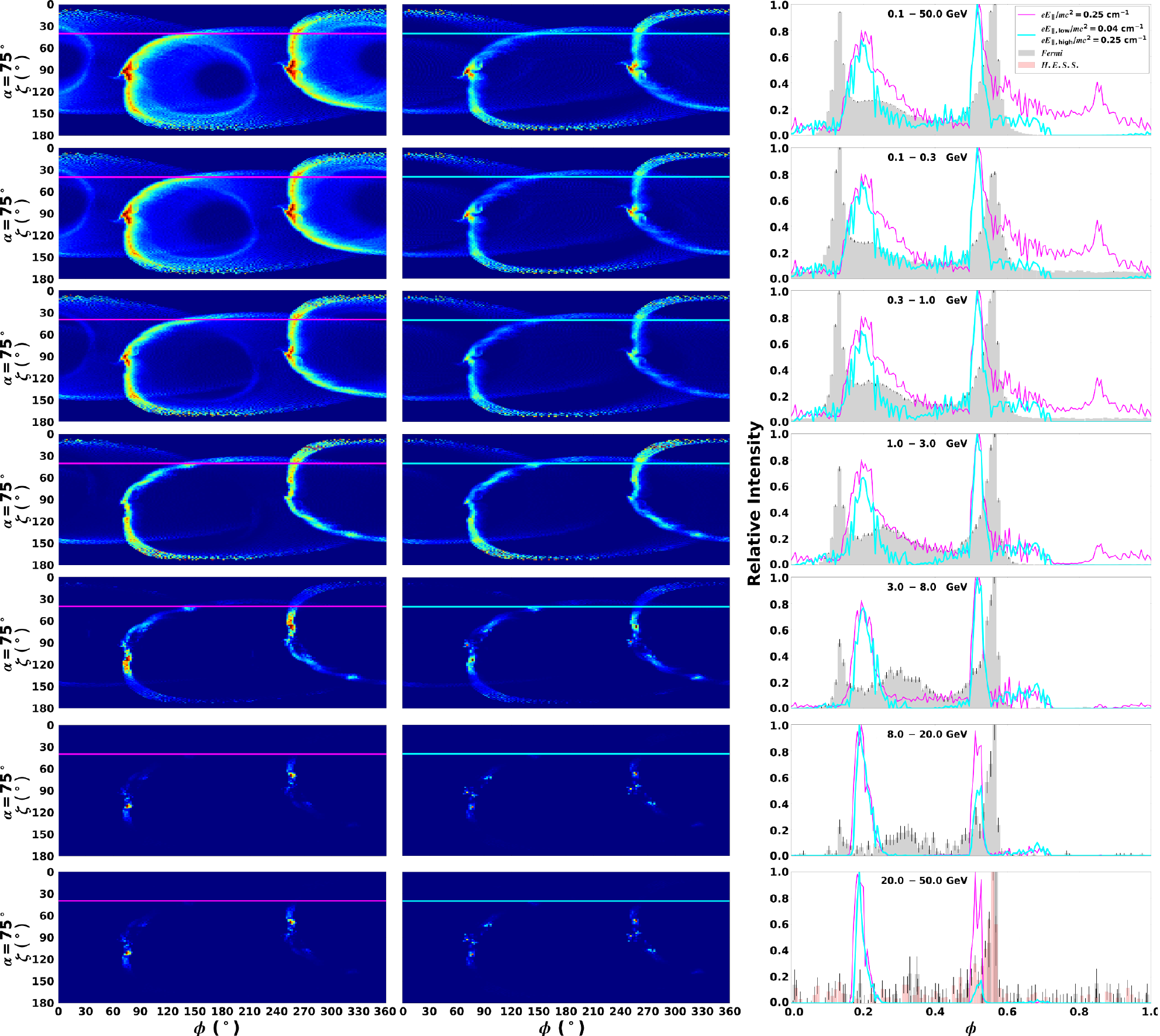}
\caption{The same as Figure~\ref{fig:PPLCopt} but for $\zeta_{\rm cut}=40^\circ$ and $\delta=-0.2$. \label{fig:PPLCrev}}
\end{figure*}

To test the robustness of the P1/P2 vs.\ $E_\gamma$ effect, we studied the light curves at $\zeta_{\rm cut}=40^\circ$ to obtain a counterexample. These light curves have a different emission structure than those in Figure~\ref{fig:PPLCopt}, due to a different spatial origin of the emission. In Figure~\ref{fig:PPLCrev} the observer misses the bridge emission because emission radiated at $\zeta_{\rm cut}=40^\circ$ is farther from the PCs than emission at $\zeta_{\rm cut}=65^\circ$. The phase plots for scenario~1 remain brighter than those for the second scenario. As the energy increases, the relative flux of P1 becomes larger than that of P2. A similar study was done by \citet{Brambilla2015} assuming a FIDO model to show that the P1/P2 effect is common, but not universal as a change in geometry can reverse the effect. Figure~\ref{fig:PPLCrev} supports this finding. We shifted the model light curves by $-0.2$ in phase to fit the \emph{Fermi} and H.E.S.S.\ data. This indicates the effect of $\zeta$ on the degeneracy of $\phi=0$ in the data (reflecting the main radio peak) and $\phi=0$ (the phase of the magnetic axis).

\begin{figure}
\centering
\epsscale{1.2}
\plotone{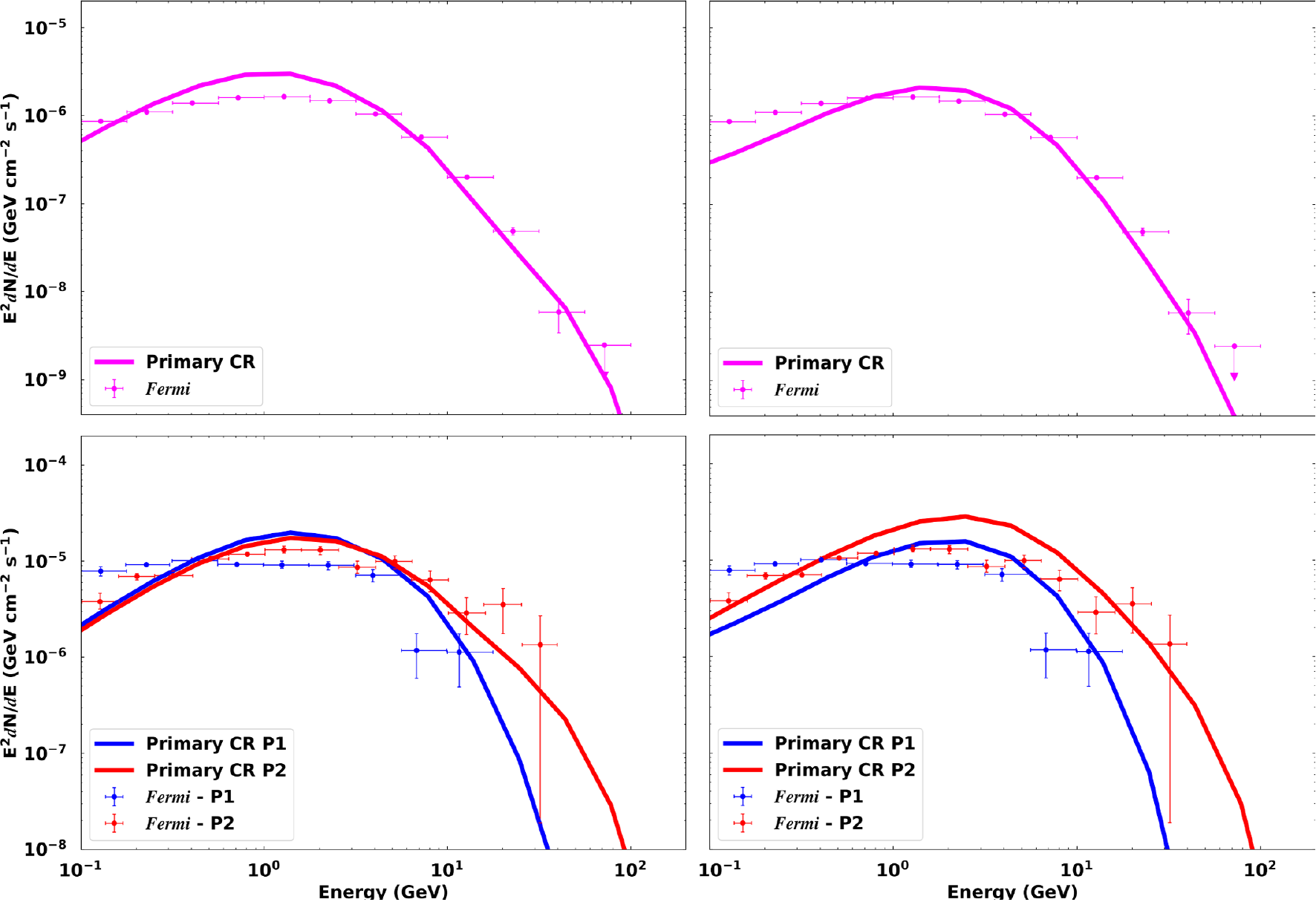}
\caption{Phase-averaged (top panel) and phase-resolved (bottom panel) spectra for the refined $\rho_{\rm c}$ calculation, for $\alpha=60^\circ$ and $\zeta_{\rm cut}=65^\circ$. For the first scenario (left column), the flux is normalized using $2J_{\rm GJ}$ and for the second case (right column), it is normalized using $3.5J_{\rm GJ}$. The data points for the phase-average spectra are from \citet{Abdo2013} (see \url{http://fermi.gsfc.nasa.gov/ssc/data/access/lat/2nd_PSR_catalog/}), and the phase-resolved spectra are updated data are from \citet{Abdo2010Vela}. \label{fig:phavgspec60}}
\end{figure}

\begin{figure}
\centering
\epsscale{1.2}
\plotone{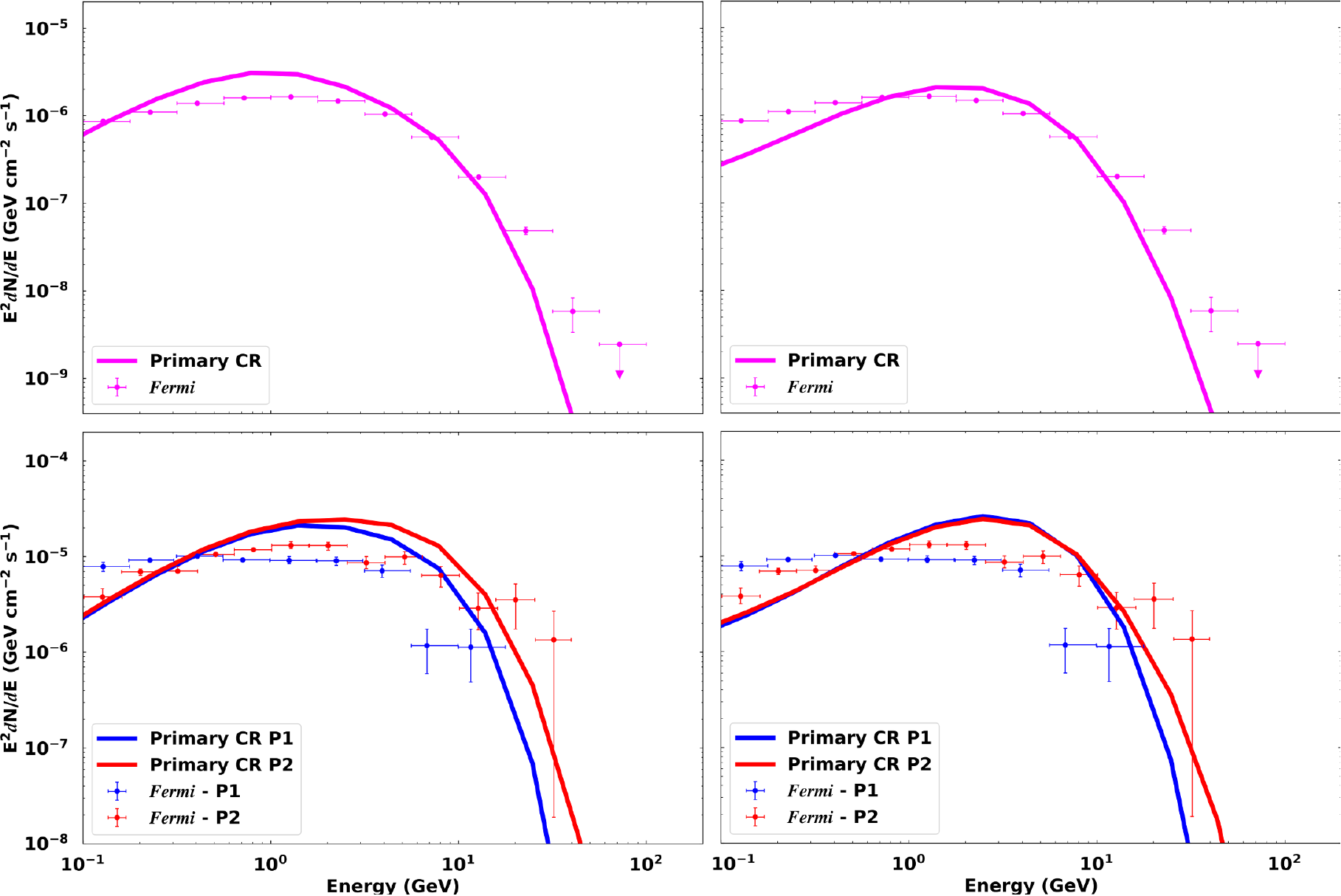}
\caption{The same as Figure~\ref{fig:phavgspec60}, but for $\alpha=75^\circ$ and $\zeta_{\rm cut}=65^\circ$. For the first scenario (left column), the flux normalization factor is $5J_{\rm GJ}$ and for the second scenario (right column), it is $8.5J_{\rm GJ}$. \label{fig:phavgspec75}}
\end{figure}

In Figure~\ref{fig:phavgspec60}, the phase-averaged and phase-resolved spectra are shown for $\alpha=60^\circ$ and $\zeta_{\rm cut}=65^\circ$. The model spectra fit the \emph{Fermi} LAT points for both $R_{\rm acc}$ cases fairly well. In the second scenario, the phase-resolved spectra of P2 have a relatively higher flux and its HE tail extends to higher $E_{\gamma}$, thus indicating a slightly larger $E_{\gamma,\rm CR}$. In Figure~\ref{fig:phavgspec75}, the phase-averaged and phase-resolved spectra are shown for the optimal parameters. The model spectra fit the data for both scenarios fairly well. In the first and second scenarios, the phase-resolved spectra of P1 and P2 are roughly equal in flux. For P2, the high-$E_{\gamma}$ tail does not extend as far as in Figure~\ref{fig:phavgspec60}, but $E_{\gamma,\rm CR}$ remains larger for P2, with the predicted cutoff being a few GeV. A larger cutoff for P2 compared to P1 is expected for this $\zeta_{\rm cut}$ value because the second light curve peak survives longer than P1 as $E_\gamma$ increases (see Figure~\ref{fig:LC_epardep}). This may not always be the case, as pointed out in Figure~\ref{fig:LC_zdep} where the P1 remains larger than P2 depending on the choice of $\zeta_{\rm cut}$.

\subsection{Effect of Refinement of the Curvature Radius} \label{subsec:refrho}
\begin{figure*}
\centering
\includegraphics[scale=0.5]{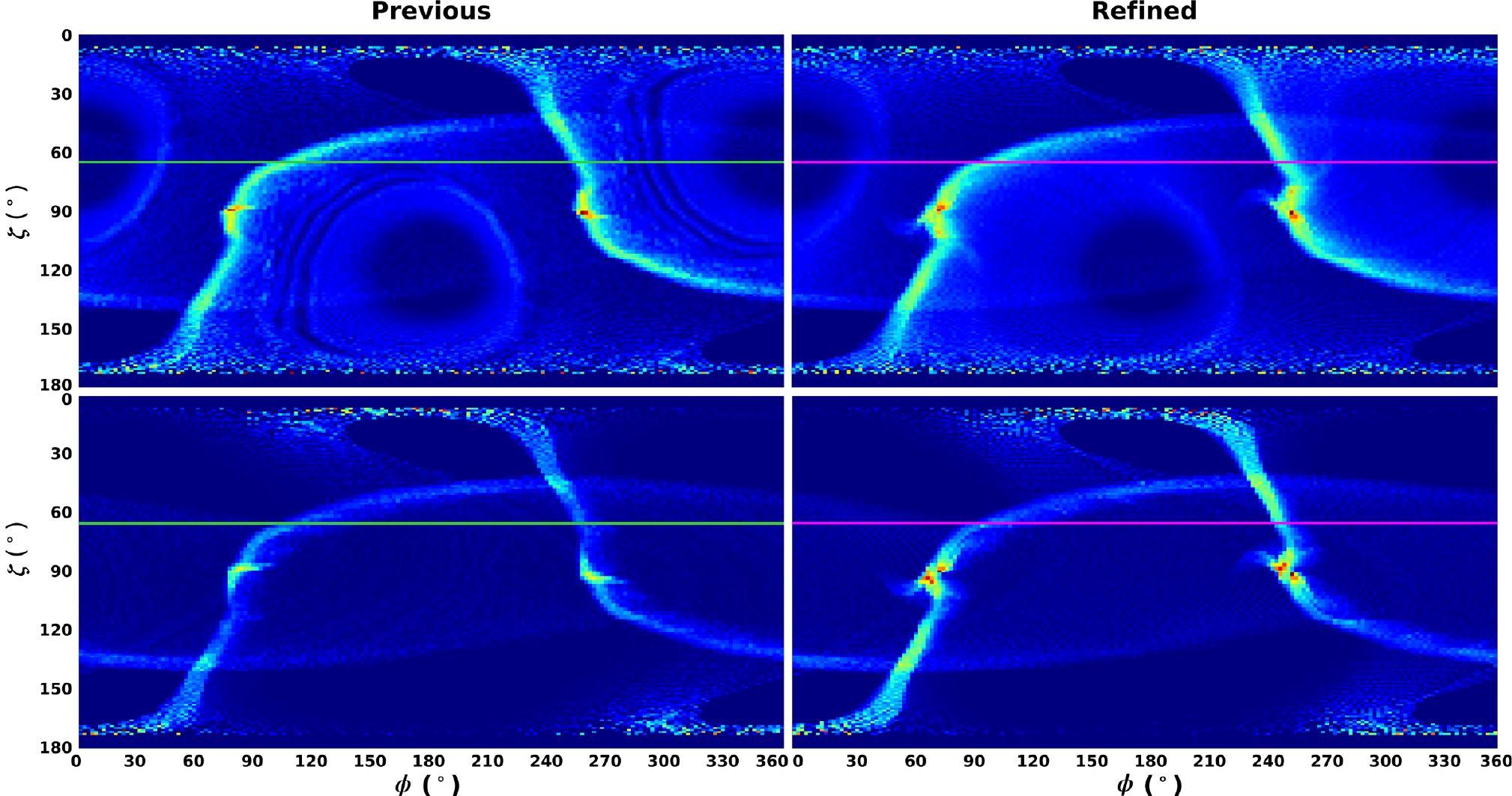}
\includegraphics[scale=0.505]{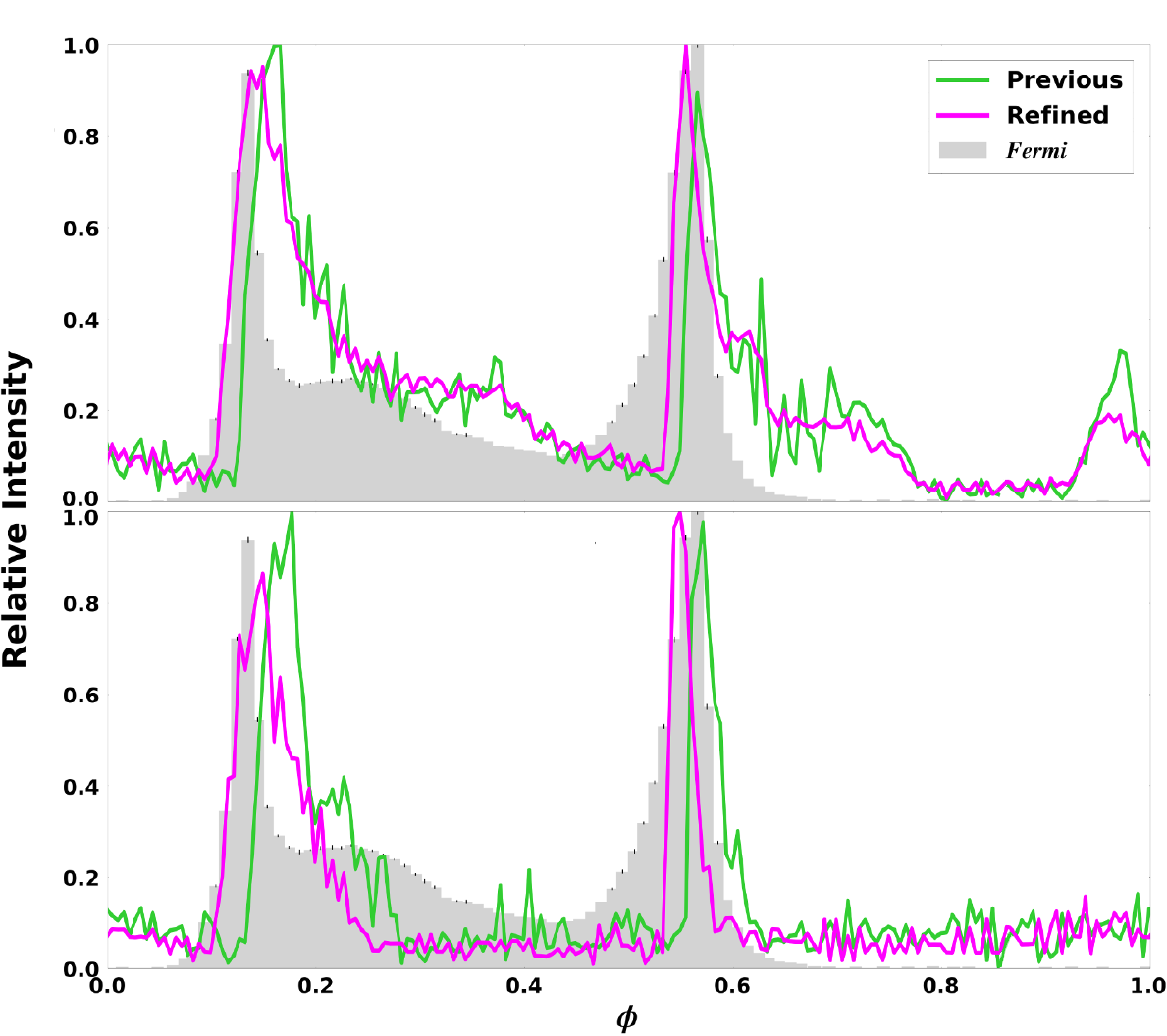}
\caption{Phase plots and pulse profiles for $\alpha=60^\circ$, $\zeta_{\rm cut}=65^\circ$, and $0.1<E_{\gamma}<50.0$~GeV. This figure serves as a comparison between phase plots for the previous (left column) and refined (center column) $\rho_{\rm c}$ calculation, and their associated light curves (right column). The top row is for a constant $E_\parallel$ (scenario~1), and the bottom row is for a two-step $E_\parallel$ (scenario~2). We shifted the resulting $\gamma$-ray model light curves by $\delta = -0.14$ in normalized phase to fit the \emph{Fermi} LAT (\citealt{Abdo2010Vela,Abdo2013},\url{http://fermi.gsfc.nasa.gov/ssc/data/access/lat/2nd_PSR_catalog/}) data points. We note that the improved trajectory calculation results in a slight shift of the light curves toward later phases. \label{fig:PPLCs_OvsN60}}
\end{figure*}
\begin{figure*}
\centering
\includegraphics[scale=0.5]{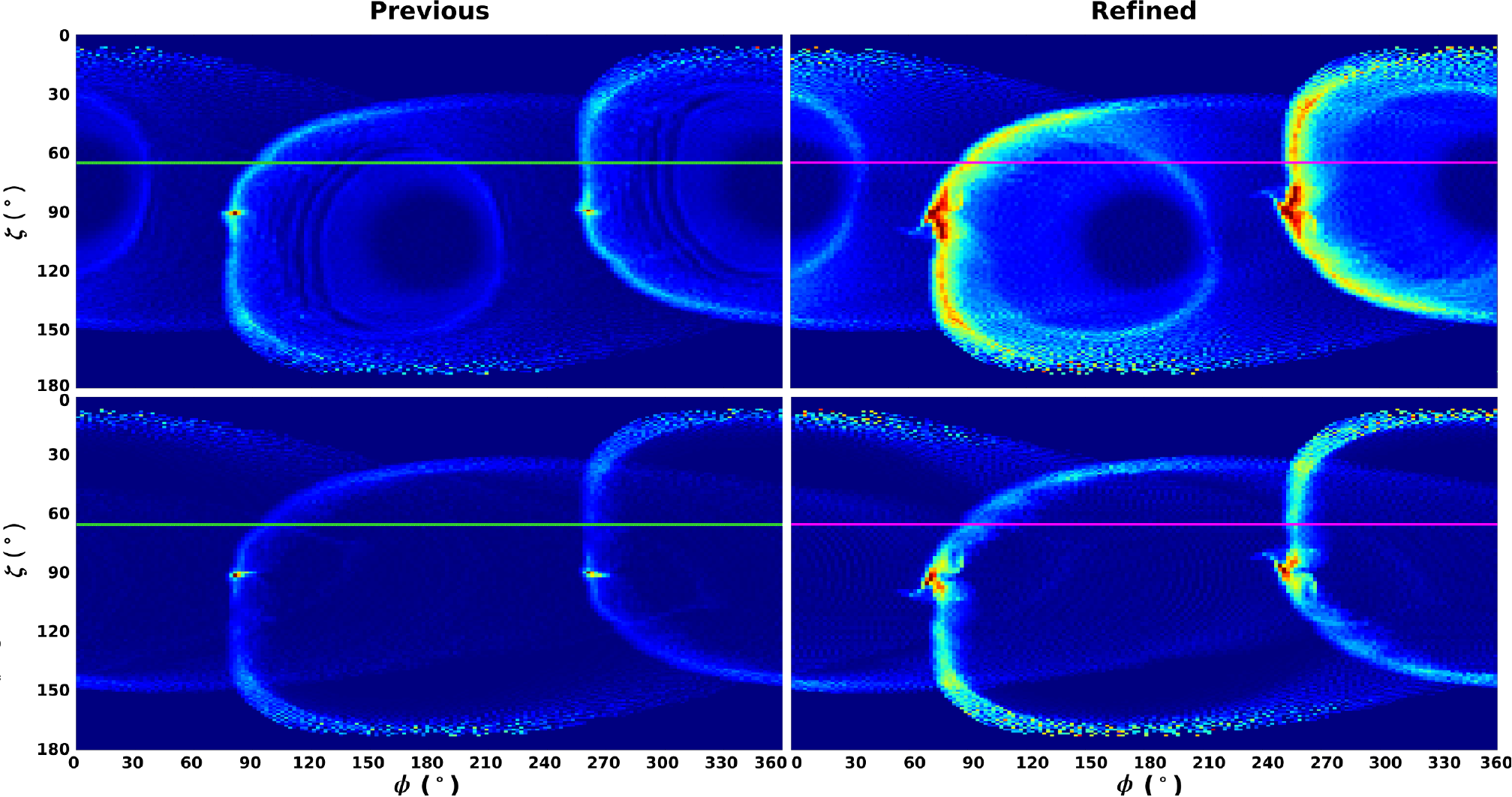}
\includegraphics[scale=0.505]{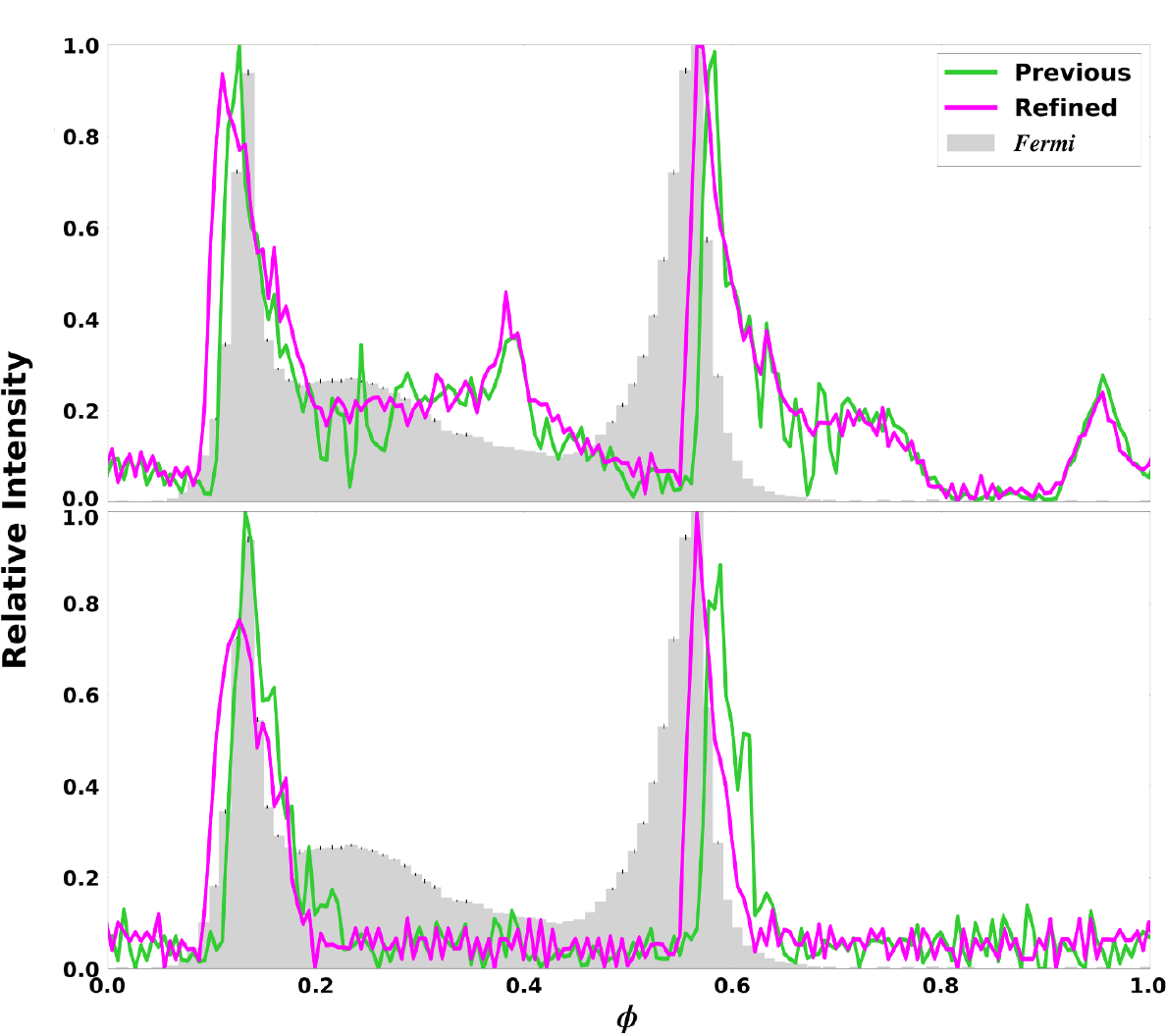}
\caption{Same as Figure~\ref{fig:PPLCs_OvsN60}, but for $\alpha=75^\circ$. \label{fig:PPLCs_OvsN75}}
\end{figure*}

In this section, we present phase plots and light curves for $\alpha=60^\circ$ and $\alpha=75^\circ$, assuming $\zeta_{\rm cut}=65^\circ$ to illustrate the effect of the previous and refined $\rho_{\rm c}$ calculation on the predicted phase plots and light curves. In Figure~\ref{fig:PPLCs_OvsN60} and \ref{fig:PPLCs_OvsN75} the phase plots and light curves associated with the previous and refined $\rho_{\rm c}$ are compared for both scenarios of $R_{\rm acc}$, and for two different values of $\alpha$. For scenario~1 (top panels) interpeak bridge emission appears at lower energies near the PCs (dark circles). This is not the case for scenario~2 (bottom panels), because $R_{\rm acc,\rm low}$ is too low at altitudes inside $R_{\rm LC}$, resulting in the suppression of the emission as well as lowering the first peak's intensity. The caustics on the phase plots for the refined $\rho_{\rm c}$ calculation, regardless of our choice of $\alpha$, appear smoother and brighter than for the previous $\rho_{\rm c}$ calculation, although their shape is largely maintained between the two calculations. A small, additional feature becomes visible near the emission caustic (indicated by the red color) of the emission when using the refined calculation. The caustics are also generally wider, and they are fuller (more filled out with radiation). The caustic shape furthermore depends strongly on the choice of $\alpha$. For $\alpha=60^\circ$ the caustic is more spread out in an S-curve shape, whereas for $\alpha=75^\circ$ it is rounded and concentrated around the PCs. The respective light curves for the two calculations are very similar, although they tend to be smoother for the refined calculation. The model light curves appear later in phase than the data, and therefore, we shifted the model with $\delta=-0.14$ in phase to fit the \emph{Fermi} data.

\subsection{Testing the Attainment of the CR Reaction Limit} \label{subsec:crr}
We solved the transport equation of a particle as it moves along a $B$-field line, focusing on CR (e.g.,~\citealt{Daugherty1982,Harding2005}; see Eq.~[\ref{eq:CRloss}]):
\begin{equation} \label{eq:transport}
{\dot{\gamma}}={\dot{\gamma}_{\rm gain}}+{\dot{\gamma}_{\rm loss}}=\frac{1}{m_{\rm e}c^2}\left[{ceE_\parallel}-\frac{2ce^2\gamma^4}{3\rho^2_c} \right],
\end{equation}
with $\dot{\gamma}$ the time derivative of $\gamma$, $\dot{\gamma}_{\rm gain}$ the acceleration rate, and $\dot{\gamma}_{\rm loss}$ the loss rate. From Eq.~(\ref{eq:transport}), it is clear that the $\dot{\gamma}_{\rm gain}$ is dependent on $R_{\rm acc}$, and $\dot{\gamma}_{\rm loss}$ is directly proportional to $\gamma^4$ and $\rho_{\rm c}^{-2}$. Eq.~(\ref{eq:transport}) may be recast in spatial terms by dividing by $c$ (assuming relativistic outflow of particles) and assuming that CR losses are dominant:
\begin{equation} \label{eq:transport1}
{\frac{d\gamma}{dl}}=R_{\rm acc}-\frac{2e^2\gamma^4}{3m_{\rm e}c^2\rho^2_c}.
\end{equation}

The CR spectral energy cutoff is defined as follows \citep[e.g.,][]{Daugherty1982,Cheng1996}:
\begin{eqnarray} \label{eq:Ecut}
E_{\gamma,{\rm CR}} = \frac{3\lambdabar_{\rm c}\gamma^3}{2\rho_{\rm c}}m_{\rm e}c^2,
\end{eqnarray}
where ${\lambdabar}_{\rm c}=\hbar/(m_{e}c)$ is the Compton wavelength and $\hbar$ the reduced Planck's constant. The curvature-radiation-reaction (CRR) limit is attained when the acceleration rate equals the loss rate. In this limit, the Lorentz factor is \citep[e.g.,][]{Luo2000}
\begin{equation}
    \gamma_{\rm CRR} = \left(\frac{3E_{\parallel}\rho_c^2}{2e}\right)^{1/4}.\label{Gamma_CRR}
\end{equation}
Substituting Eq.~(\ref{Gamma_CRR}) into Eq.~(\ref{eq:Ecut}), we obtain for a constant $E$-field \citep{Venter2010}
\begin{eqnarray} \label{eq:EcutCRR}
  E_{\gamma,{\rm CR}} \sim 4\left(\frac{E_{\parallel}}{10^4~{\rm statvolt\,cm^{-1}}}\right)^{3/4}\left(\frac{\rho_{c}}{10^8~{\rm cm}}\right)^{1/2} 
\end{eqnarray}
measured in GeV. We generally test if the CRR limit is attained in both scenarios by plotting the $\log_{10}$ of $\rho_{\rm c}$, $\gamma$, $\dot{\gamma}_{\rm gain}$, and $\dot{\gamma}_{\rm loss})$ along the same field lines chosen in Figure~\ref{fig:posdirrhoOvsN}, checking if the acceleration and loss rates become equal at large distances. The particle dynamics depend on the $\rho_{\rm c}$, therefore an improved calculation yielding a smoother $\rho_{\rm c}$ has an impact on the particle transport and thus the energy-dependent light curves and spectra. 
\begin{figure}
\epsscale{1.0}
\plotone{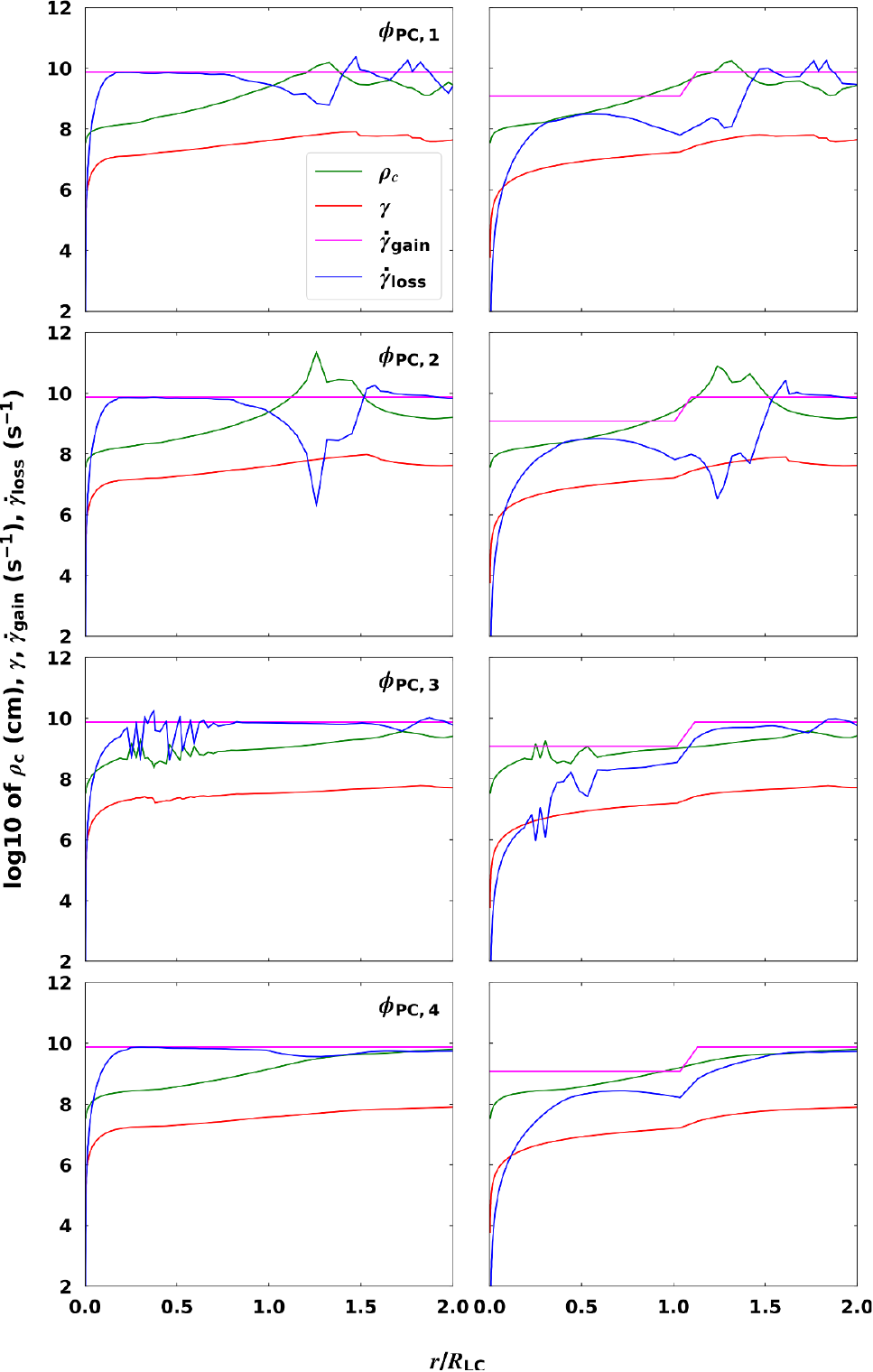}
\caption{The particle dynamics, along the same $B$-field lines as in Figure~\ref{fig:posdirrhoOvsN}, are shown for the refined $\rho_{\rm c}$ calculation, for $\alpha=75^\circ$. The quantities plotted are the $\log_{10}$ of $\rho_{\rm c}$ (green), $\gamma$ (red), $\dot{\gamma}_{\rm gain}$ (magenta), and $\dot{\gamma}_{\rm loss}$ (blue), for both scenario~1 (left column) and scenario~2 (right column). \label{fig:crr}}
\end{figure}

In Figure~\ref{fig:crr}, the CRR limit is almost immediately attained in the first scenario, because the $E_\parallel$ is large enough to supply the primaries with ample energy at lower altitudes. The rapid rise of $\gamma$ to $\sim5\times10^7$ leads to a rapid increase in $\dot{\gamma}_{\rm loss}$, and then the CRR limit is reached around $0.2R_{\rm LC}$. However, because $\rho_{\rm c}$ oscillates or dips along some of the field lines, this limit is disturbed (because the loss rate is anticorrelated with $\rho_{\rm c}$) and in some cases only recovered later on at higher altitudes. Indeed, instabilities in $\rho_{\rm c}$ cause similar but anti-correlated oscillations in $\dot{\gamma}_{\rm loss}$. If the $E_\parallel$ is lower inside the light cylinder, as in the second scenario, the acceleration of the primaries is initially suppressed, as is $\dot{\gamma}_{\rm loss}$. However, beyond $r\sim R_{\rm LC}$, where a higher $E_\parallel$ is assumed, the particles accelerate efficiently, and the CRR limit may be reached around $r\sim 1.5R_{\rm LC}$.

\subsection{Local Environment of Emission Regions Connected to Each Light Curve Peak} \label{subsec:environment}
In order to isolate and understand the P1/P2 vs.\ $E_\gamma$ effect seen in the light curves of Vela, we investigated the values of $E_{\gamma,\rm CR}$ (Eq.~[\ref{eq:Ecut}]), $\rho_{\rm c}$ and $\gamma$ 
in the spatial regions where each model peak originates for the set of optimal parameters we found as described in Section~\ref{subsec:optparams}.
Additionally, we considered the distribution of $r$ (radial coordinate, in units of $R_{\rm LC}$) for the emission contributing to each peak.

Thus, as explained in Section~\ref{sec:revmap}, we perform ``reverse mapping'' and accumulate the range of values that each of these quantities assume in the regions where the photons originate that make up P1 and P2, for a particular selected photon energy range. Thus, these quantities are in principle calculated for all $E_\gamma$, but we subsequently apply cuts in $E_\gamma$ and then study the resulting distributions of the mentioned quantities associated with photons in a particular chosen band. These binned quantities are therefore presented as $E_\gamma$-dependent histograms below, where we scaled the frequency of occurrence of the quantities (signifying an unweighted probability) using the emitted photon emission rate $\dot{N}_{\gamma}$, to obtain a true (weighted) relative probability for each quantity, and for the chosen energy range, as indicated on the $y$-axis.

\begin{figure*}
\gridline{\fig{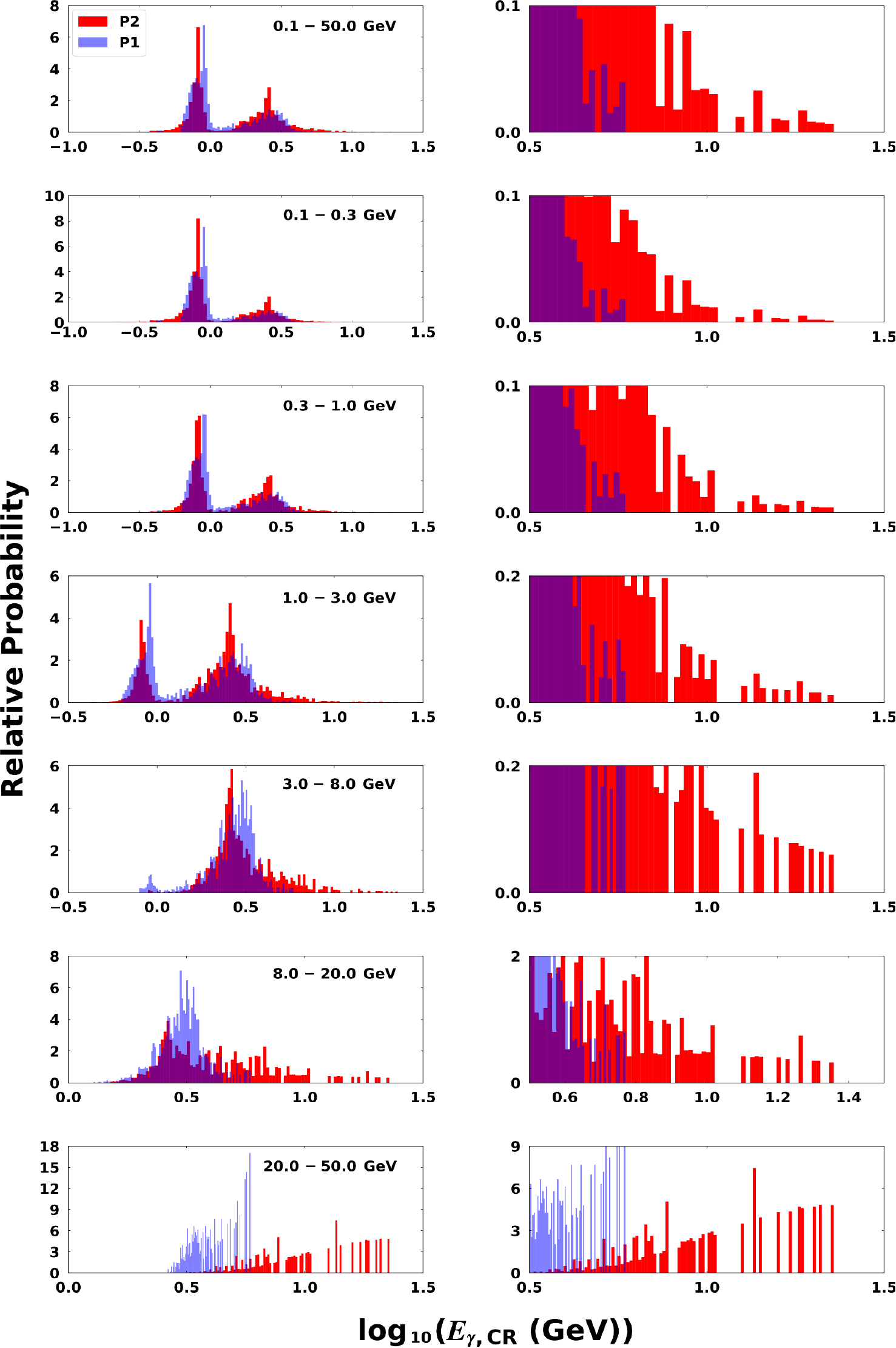}{0.45\textwidth}{(a)}
          \fig{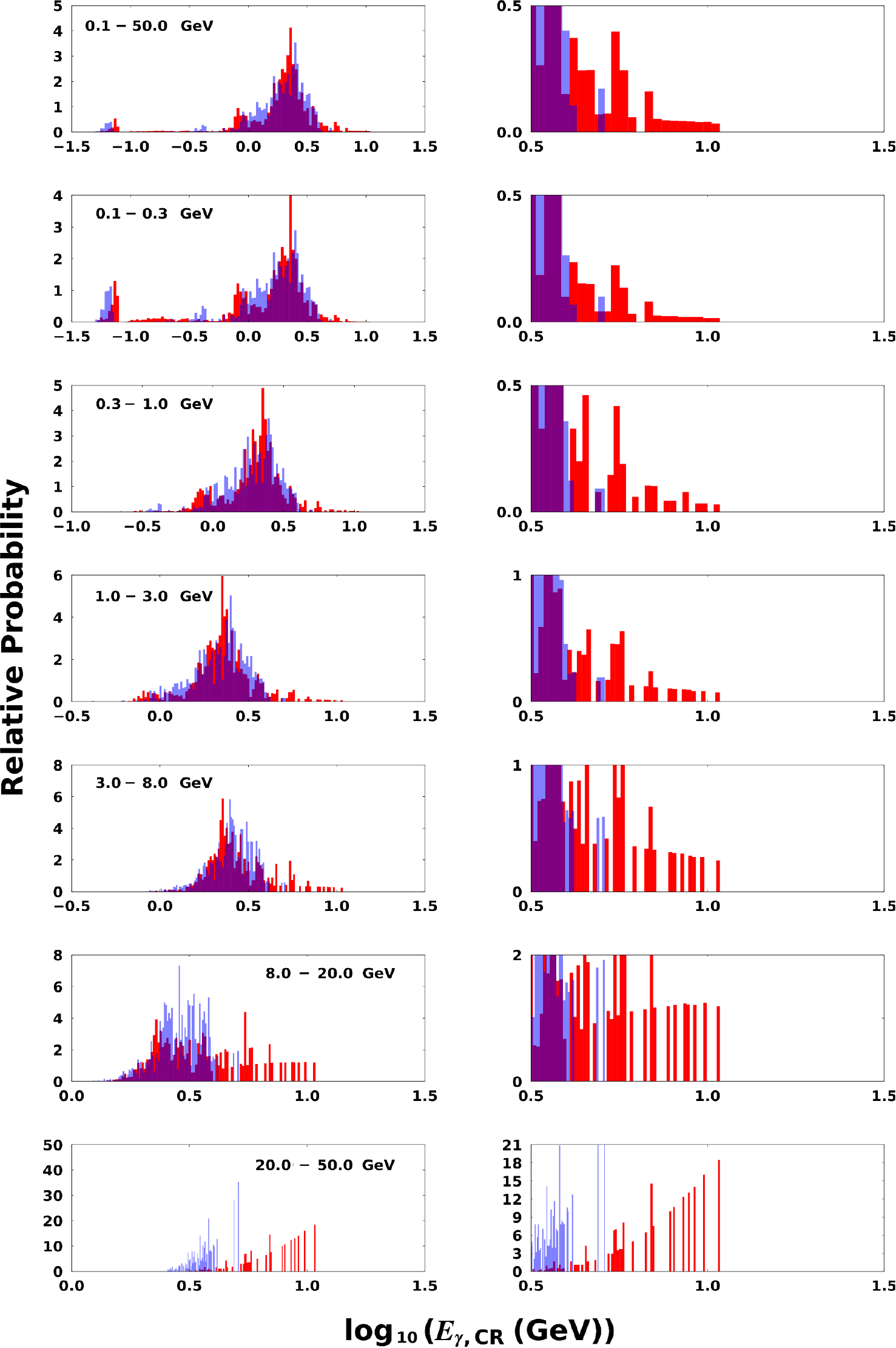}{0.45\textwidth}{(b)}
         }
\caption{Energy-dependent histograms for $\log_{10}(E_{\gamma,\rm CR}/{\rm GeV})$ (the spectral cutoff) for P1 (blue curve) and P2 (red curve), where (a) represents scenario~1 and (b) scenario~2. The respective energy bands are indicated by labels in each panel. The second column in each case represents a zoom-in of the tails of the distributions for large values of $\log_{10}(E_{\gamma,\rm CR}/{\rm GeV})$. \label{fig:hists_ecut}}
\end{figure*}

In Figure~\ref{fig:hists_ecut} we show histograms for $\log_{10}(E_{\gamma,\rm CR}/{\rm GeV})$, for different photon energy ranges. This quantity is calculated using Eq.~(\ref{eq:Ecut}), specifically involving $\rho_{\rm c}$ and $\gamma$. We calculate this $E_{\gamma,\rm CR}$ histogram for different $E_{\gamma}$ ranges as follows. We follow the motion of particles along their trajectories. For a fixed photon energy, we calculate $E_{\gamma,\rm CR}$ at each spatial position of the particles, because they radiate a spectrum of $\gamma$-ray photons and then bin these cutoff energies. We use the fixed photon energies to define the ranges (as noted in the labels of Figure~\ref{fig:hists_ecut}). It may therefore happen that a photon with a low chosen energy forms part of a CR spectrum only cutting off at a much larger energy, or vice versa if the photon originates in the tail of the single-particle CR spectrum. For example, one finds that the $E_{\gamma,\rm CR}$ histogram for P2 has a tail extending out to $\sim20$~GeV, even though the chosen photon energy cut is $0.1-1.0$~GeV (third row, first column). Thus, the spectral cutoff exceeds the maximum photon energy of the chosen range. Conversely (last row, first column), $E_{\gamma,\rm CR}$ is below the maximum chosen photon energy of 50~GeV, because the spectrum cuts off well below this maximum photon energy. The cutoff energy is thus not a one-to-one function of photon energy, since a single-particle CR spectrum includes many photon energies but only a single spectral cutoff. Yet, by cutting on photon energy, one can isolate low or high energies and probe the value for $E_{\gamma,\rm CR}$, $\rho_{\rm c}$ and $\gamma$ with photon energy. It is thus important to scale the occurrence of these quantities with the number of photons at a particular fixed photon energy and for a given step length at a particular point along the particle trajectories.

In the first scenario (left column of Figure~\ref{fig:hists_ecut}), there appears two bumps, for both peaks, at lower $E_\gamma$ (up to $\sim5$~GeV), situated around $\log_{10}(E_{\gamma,\rm CR}/{\rm GeV})\approx{-0.2}$ and ${0.4}$. The lower bump disappears with increasing $E_\gamma$. 
In (b), we show scenario~2 where there is a small low-$E_\gamma$ bump (up to $\sim0.3$~GeV) at even smaller values of $\log_{10}(E_{\gamma,\rm CR}/{\rm GeV})\approx{-1.2}$. The existence of this bump is probably because of the lower value of $R_{\rm acc}$ inside the light cylinder that suppresses the low-altitude acceleration and emission in this scenario. Also, the lower-energy bump disappears as the $E_\gamma$ is increased, because only photons from individually radiated spectra (which make up the cumulative spectrum seen by the observer) with higher cutoffs are then visible. The $\log_{10}(E_{\gamma,\rm CR}/{\rm GeV})$ of P2 is relatively larger than that of P1 for both scenarios, as seen in the zoom-ins. This confirms what has already been seen in the light curves in Figure~\ref{fig:PPLCopt} and spectra in Figure~\ref{fig:phavgspec75}: P2 survives with an increase in energy because its spectral cutoff is relatively higher than that of P1. The $\log_{10}(E_{\gamma,\rm CR}/{\rm GeV})$ of P2 reaches values as high as $\sim10^{1.0}-10^{1.4}$, with larger values reached in the first scenario, given the higher $E$-field. 
\begin{figure*}
\gridline{\fig{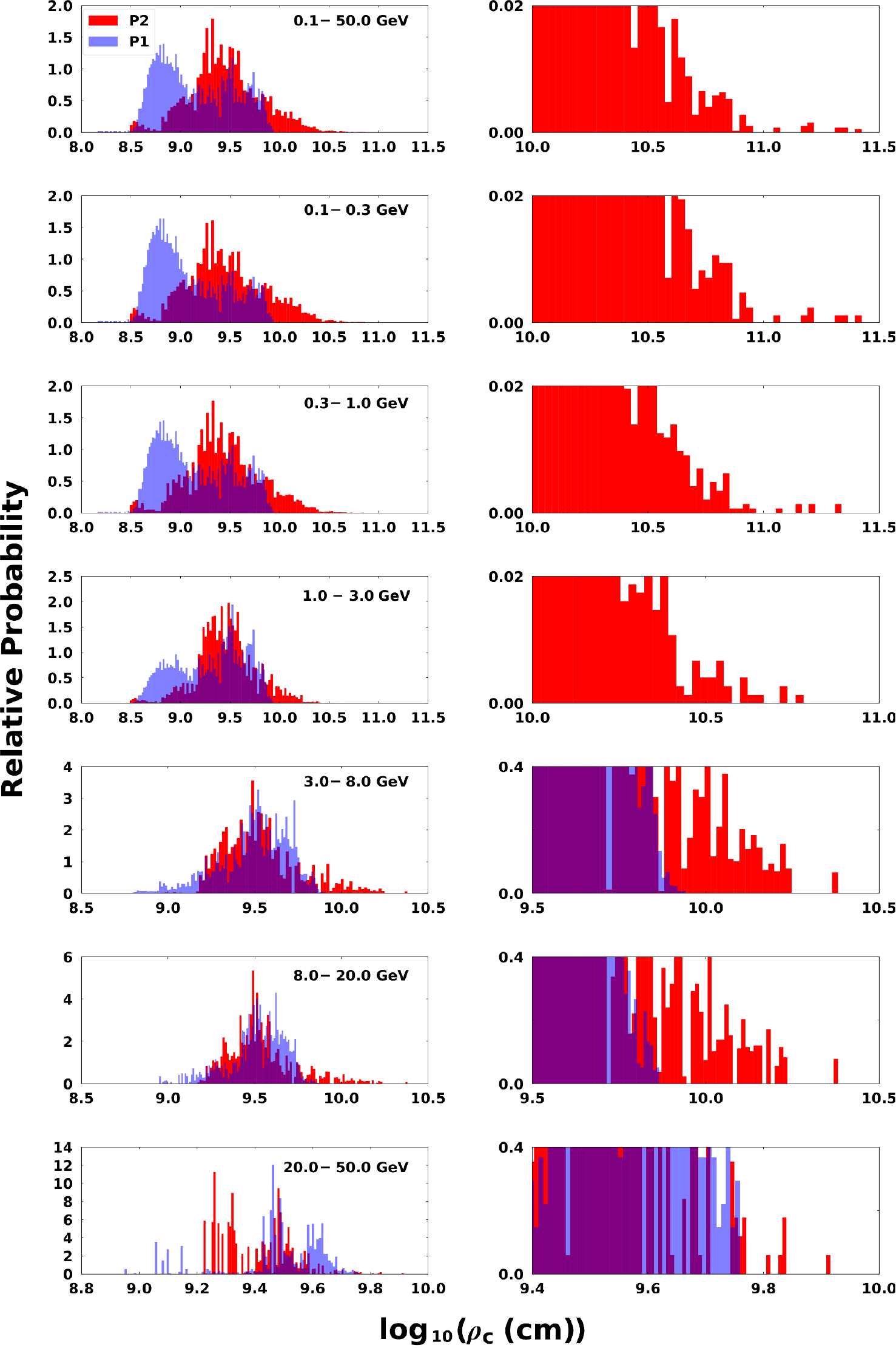}{0.45\textwidth}{(a)}
          \fig{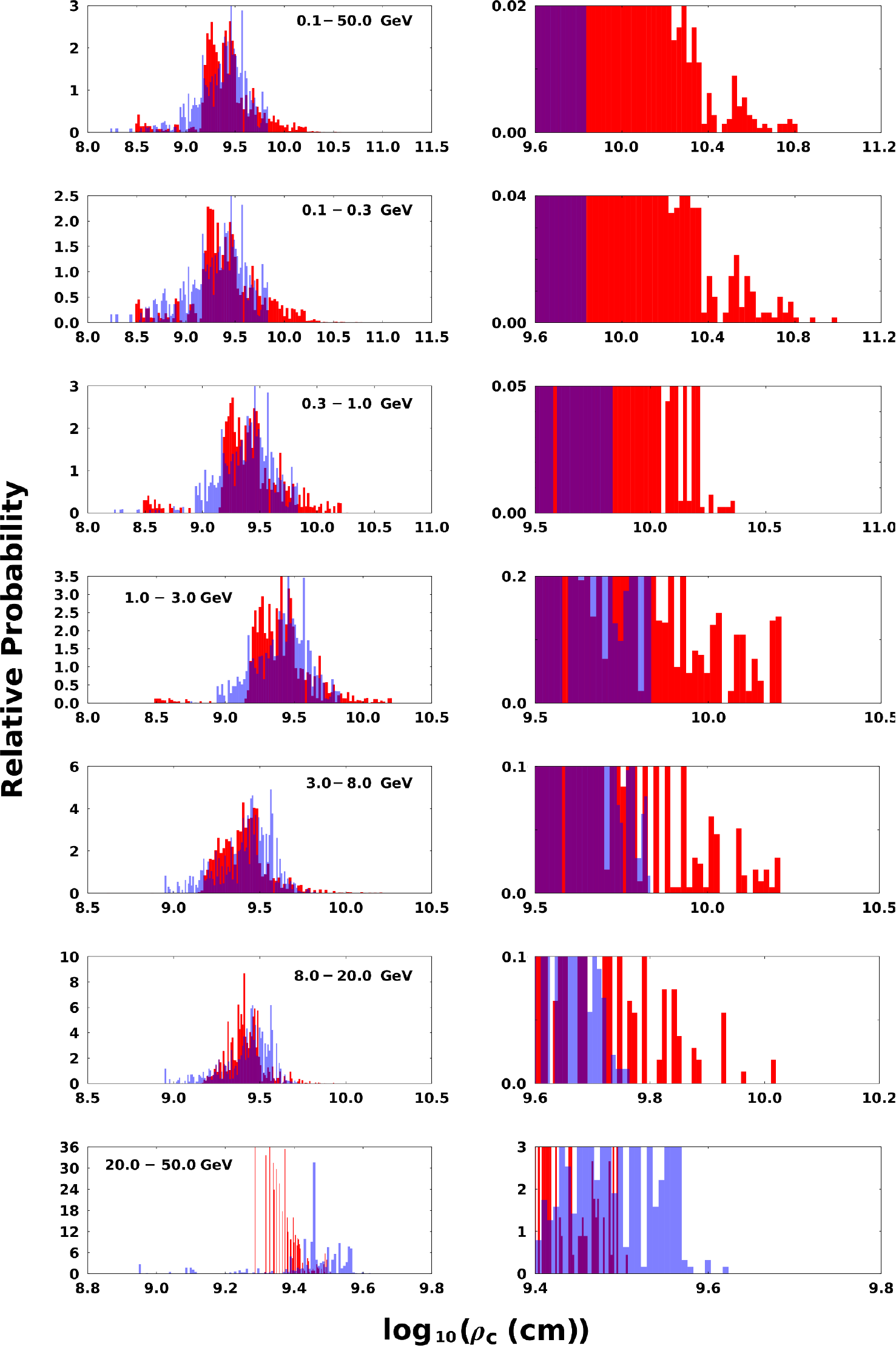}{0.45\textwidth}{(b)}
         }
\caption{The same as Figure~\ref{fig:hists_ecut} but for $\log_{10}(\rho_{\rm c})$. \label{fig:hists_rho}}
\end{figure*}

In Figure~\ref{fig:hists_rho} we show histograms of the relative probability as a function of $\log_{10}(\rho_{\rm c})$ for P1 (blue) and P2 (red). For the first scenario in (a), with a zoom-in of the tail of the distributions (right column), a bump appears around $\log_{10}(\rho_{\rm c})\approx{8.5}$ to~$9.0$ for P1 at lower $E_\gamma$ (up to $\sim3$~GeV), which disappears with increasing $E_\gamma$. In (b) we show scenario~2, where there is no low-$E_\gamma$ bump at smaller values of $\log_{10}(\rho_{\rm c})$ as in the first scenario. This is due to the fact that in the first scenario, the accelerating $E$-field is relatively larger at lower altitudes, so that the particles can radiate in the GeV band from these lower altitudes characterized by lower values of $\log_{10}(\rho_{\rm c})$. In the second scenario, however, the small value of $R_{\rm acc,\rm low}$ inside the light cylinder suppresses emission in the GeV band originating from lower altitudes, hence the missing bump. Importantly, the $\log_{10}(\rho_{\rm c})$ of P2 is relatively larger than that of P1 for both scenarios, as seen in the zoom-ins, with P2's associated $\rho_{\rm c}$ reaching values as high as $\sim10^{9.8}-10^{11.5}$~cm (indicating relatively less curved orbits). The $\rho_{\rm c}$ values reached in scenario~1 for P2 are also relatively larger than those in scenario~2 for the same peak. Thus, for sustained acceleration, particles radiating at high energies are moving along slightly straighter orbits. 
It is only at energies above 20~GeV that the values of $\log_{10}(\rho_{\rm c})$ associated with P1 become comparable to or larger than those associated with P2 in scenario~2.

\begin{figure*}
\gridline{\fig{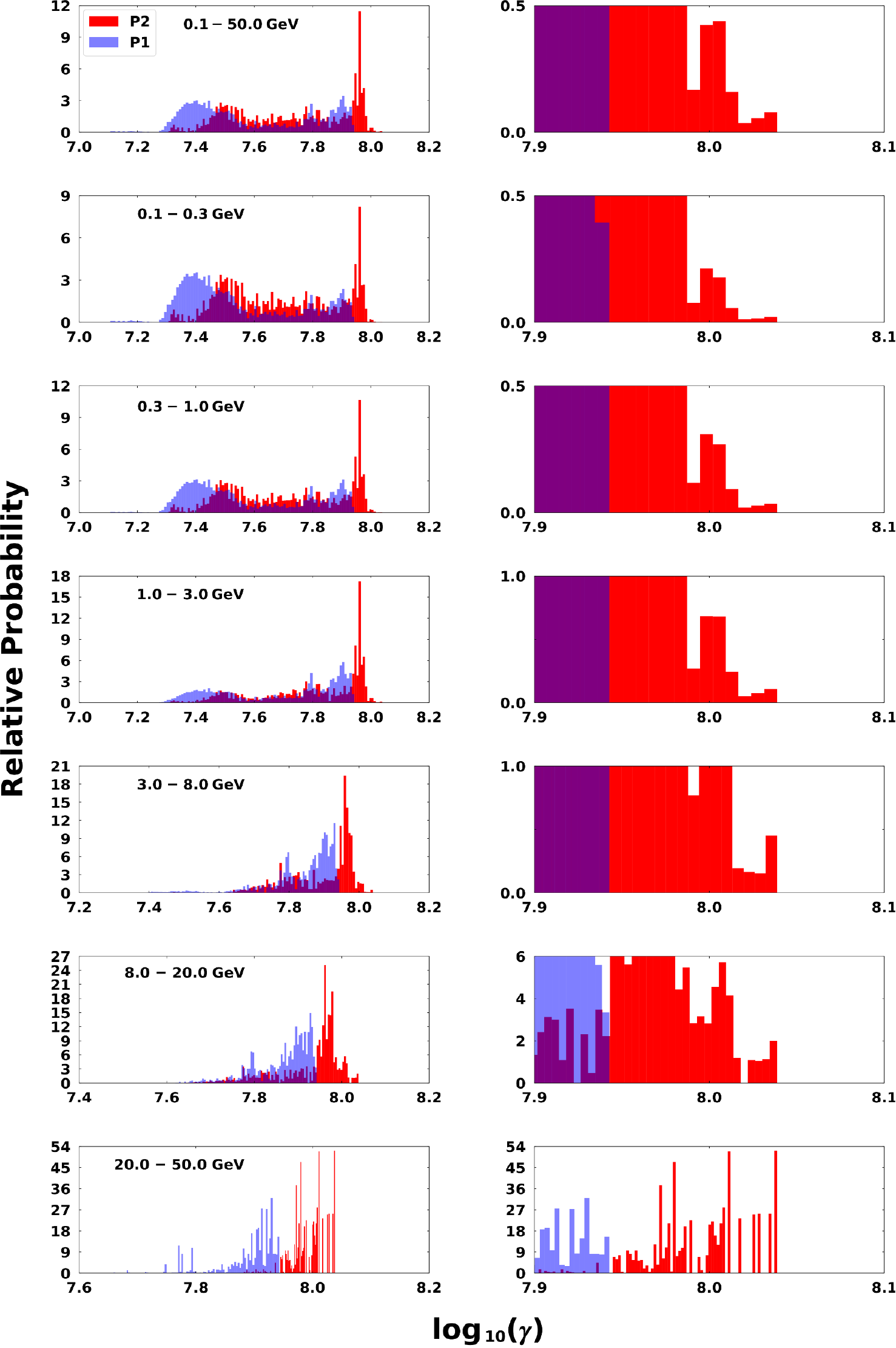}{0.45\textwidth}{(a)}
          \fig{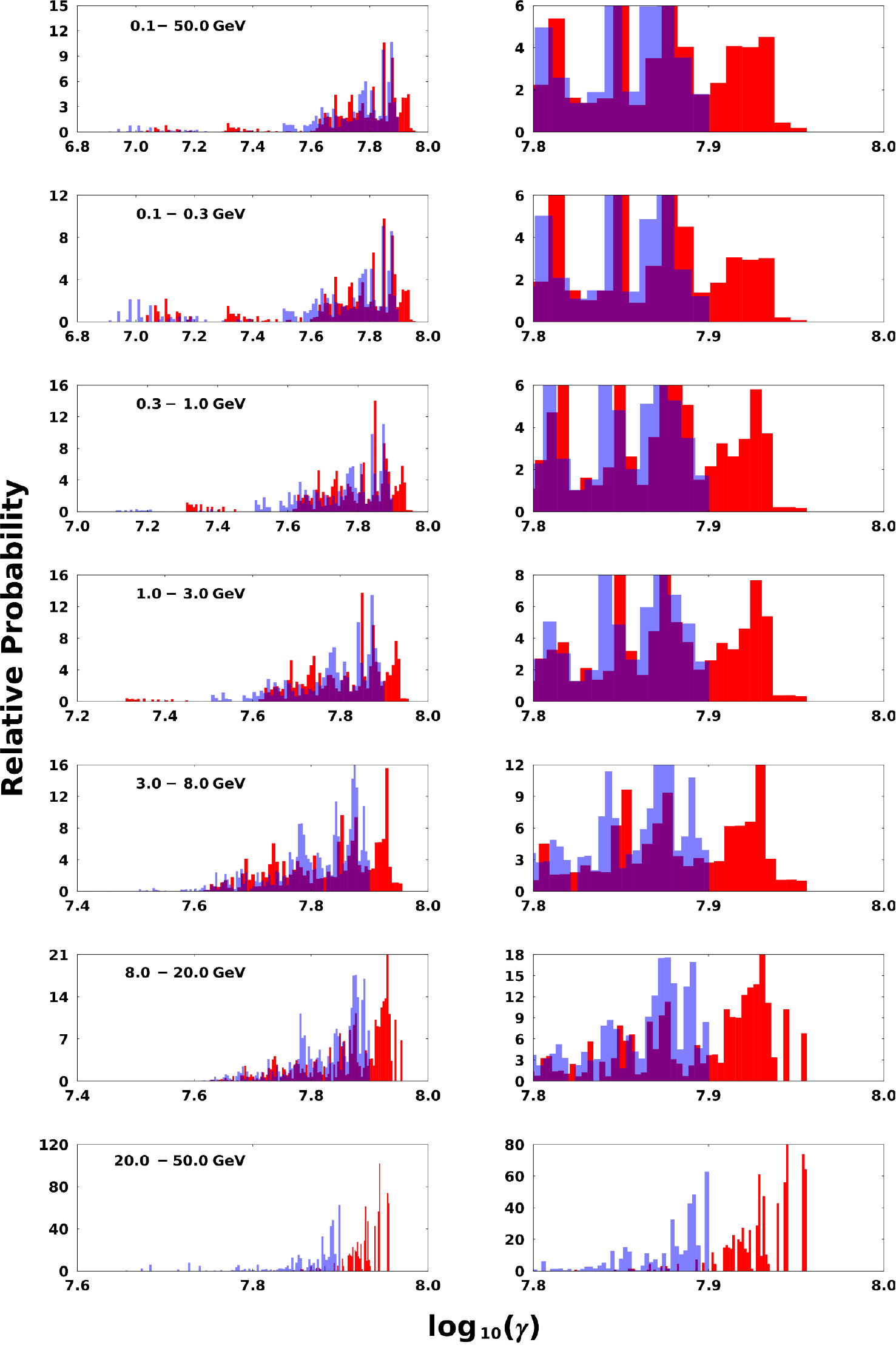}{0.45\textwidth}{(b)}
         }
\caption{The same as Figure~\ref{fig:hists_ecut} but for $\log_{10}(\gamma)$. \label{fig:hists_gam}}
\end{figure*}
Similar to Figure~\ref{fig:hists_ecut} and Figure~\ref{fig:hists_rho}, we show histograms of $\log_{10}(\gamma)$ in Figure~\ref{fig:hists_gam} for different energy ranges. In the first scenario indicated in (a), a bump appears around $\log_{10}(\gamma)\approx{7.3}-7.5$ for both peaks at lower $E_\gamma$ (up to $\sim3$~GeV), which disappears with increasing $E_\gamma$. In (b) we show scenario~2 where there is no low-$E_\gamma$ bump at smaller values of $\log_{10}(\gamma)$. There is also a peak in $\log_{10}(\gamma)\sim8$ for P2 in scenario~1, while $\log_{10}(\gamma)$ is relatively smaller in scenario~2, given the fact that particles experienced less acceleration in that case. The $\log_{10}(\gamma)$ of P2 is relatively larger than that of P1 as seen in the zoom-ins for both scenarios.

\begin{figure*}
\gridline{\fig{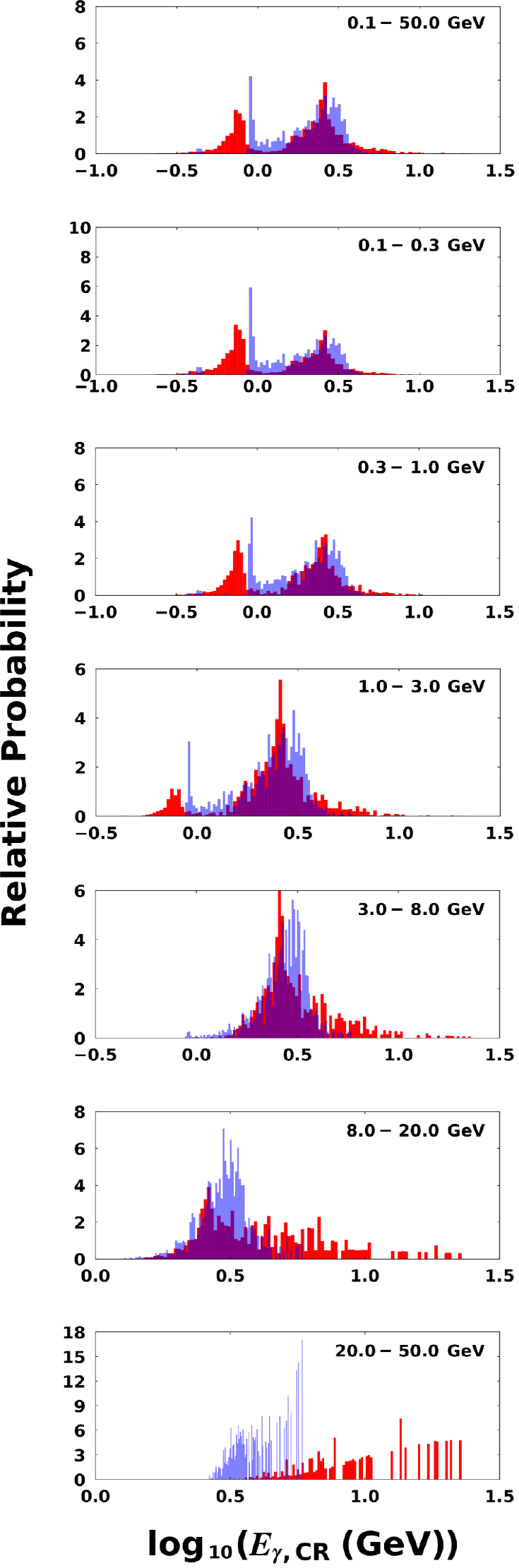}{0.27\textwidth}{(a)}
          \fig{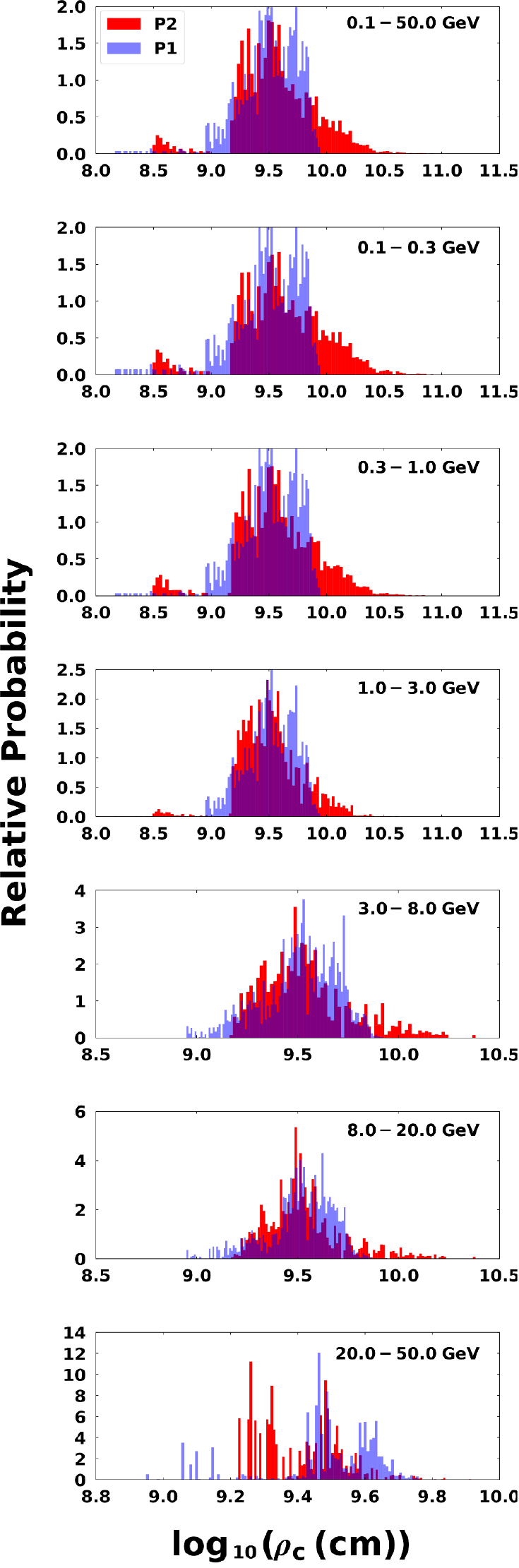}{0.268\textwidth}{(b)} 
		 \fig{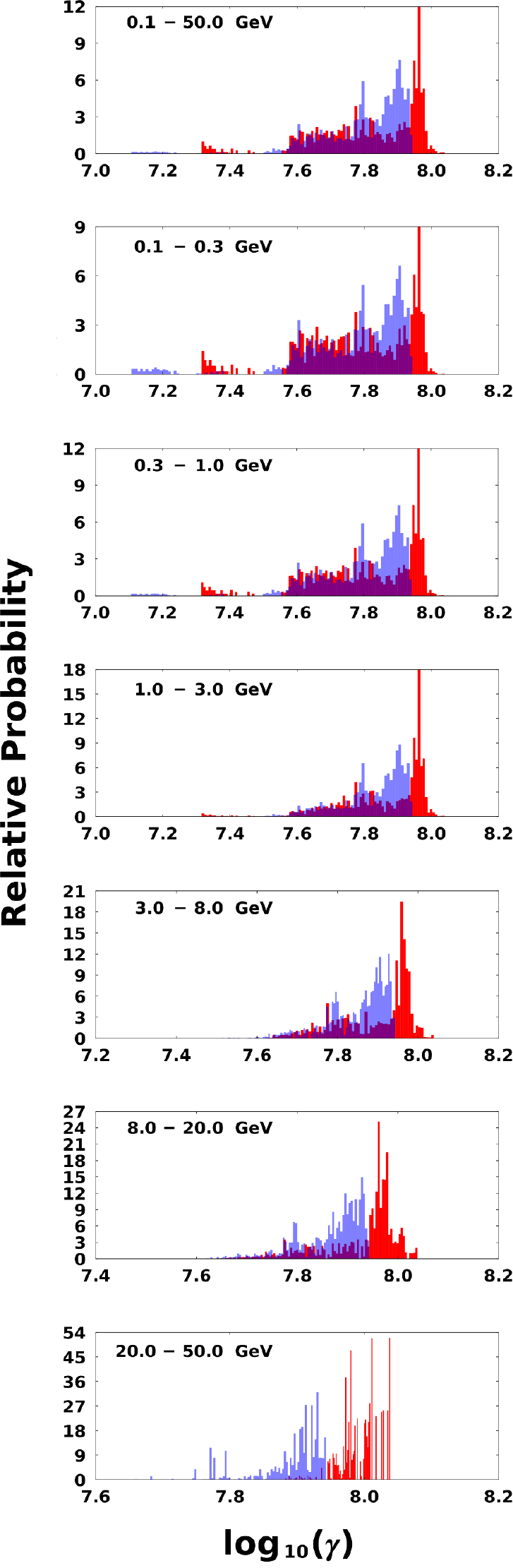}{0.27\textwidth}{(c)}
         }
\caption{Energy-dependent histograms of (a) $\log_{10}(E_{\gamma,\rm CR}/{\rm GeV})$, (b) $\log_{10}(\rho_{\rm c})$, and (c) $\log_{10}(\gamma$), for P1 (blue curve) and P2 (red curve). All three cases are for the first scenario at altitudes equal to and beyond $R_{\rm LC}$, into the current sheet. \label{fig:hists_RlcLim1}}
\end{figure*}
\begin{figure*}
\gridline{\fig{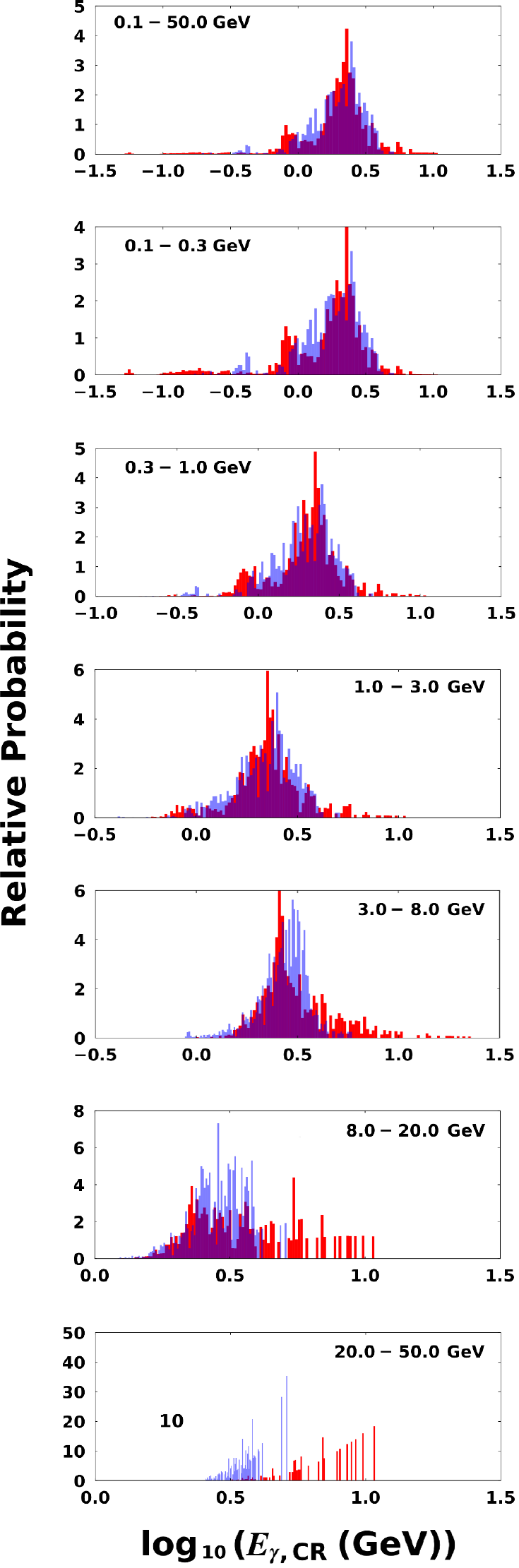}{0.27\textwidth}{(a)}
          \fig{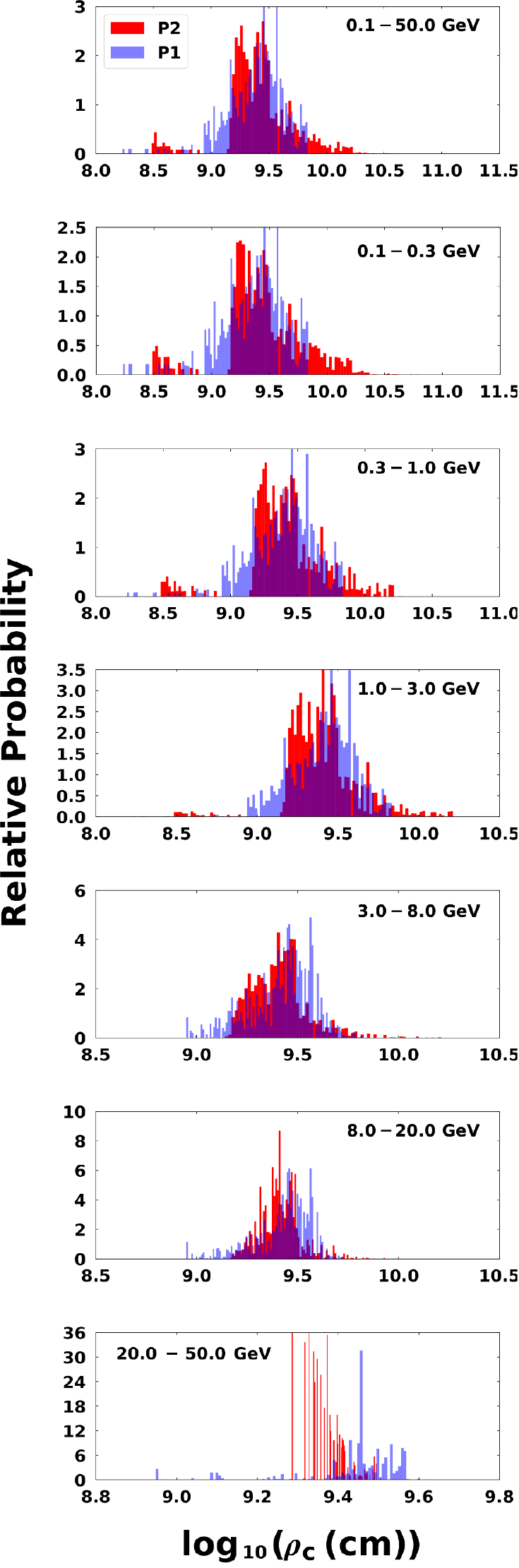}{0.268\textwidth}{(b)}
		 \fig{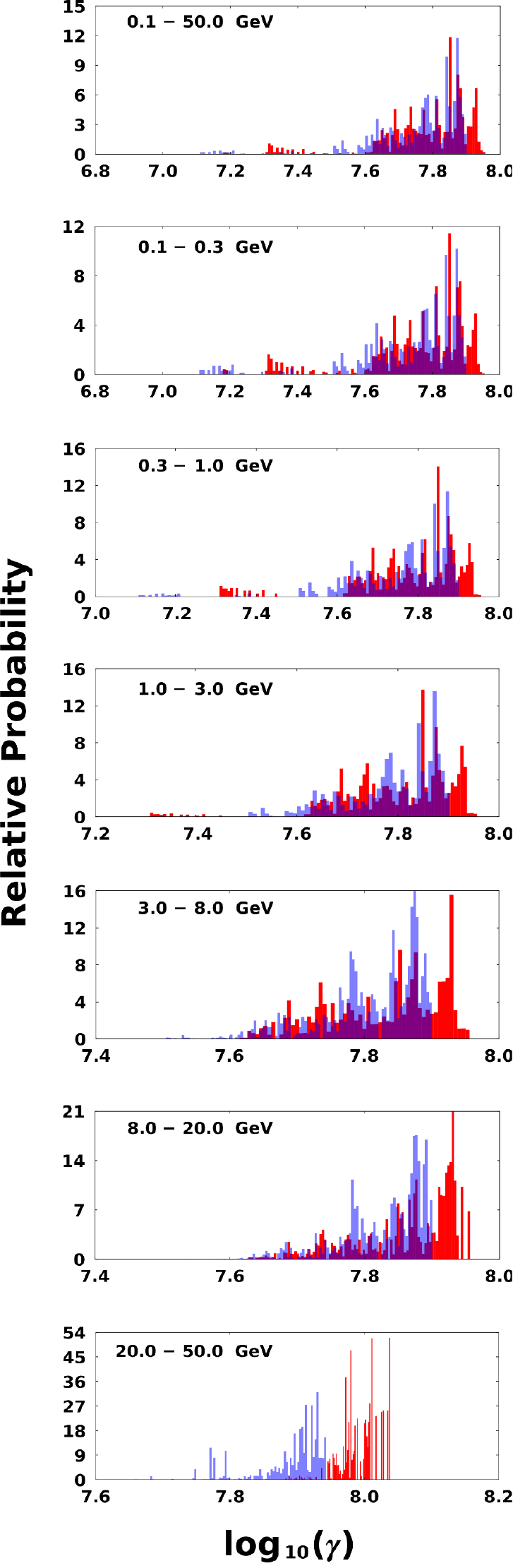}{0.27\textwidth}{(c)}
         }
\caption{The same as in Figure~\ref{fig:hists_RlcLim1} but for the second scenario. \label{fig:hists_RlcLim2}}
\end{figure*}

\begin{figure*}
\gridline{\fig{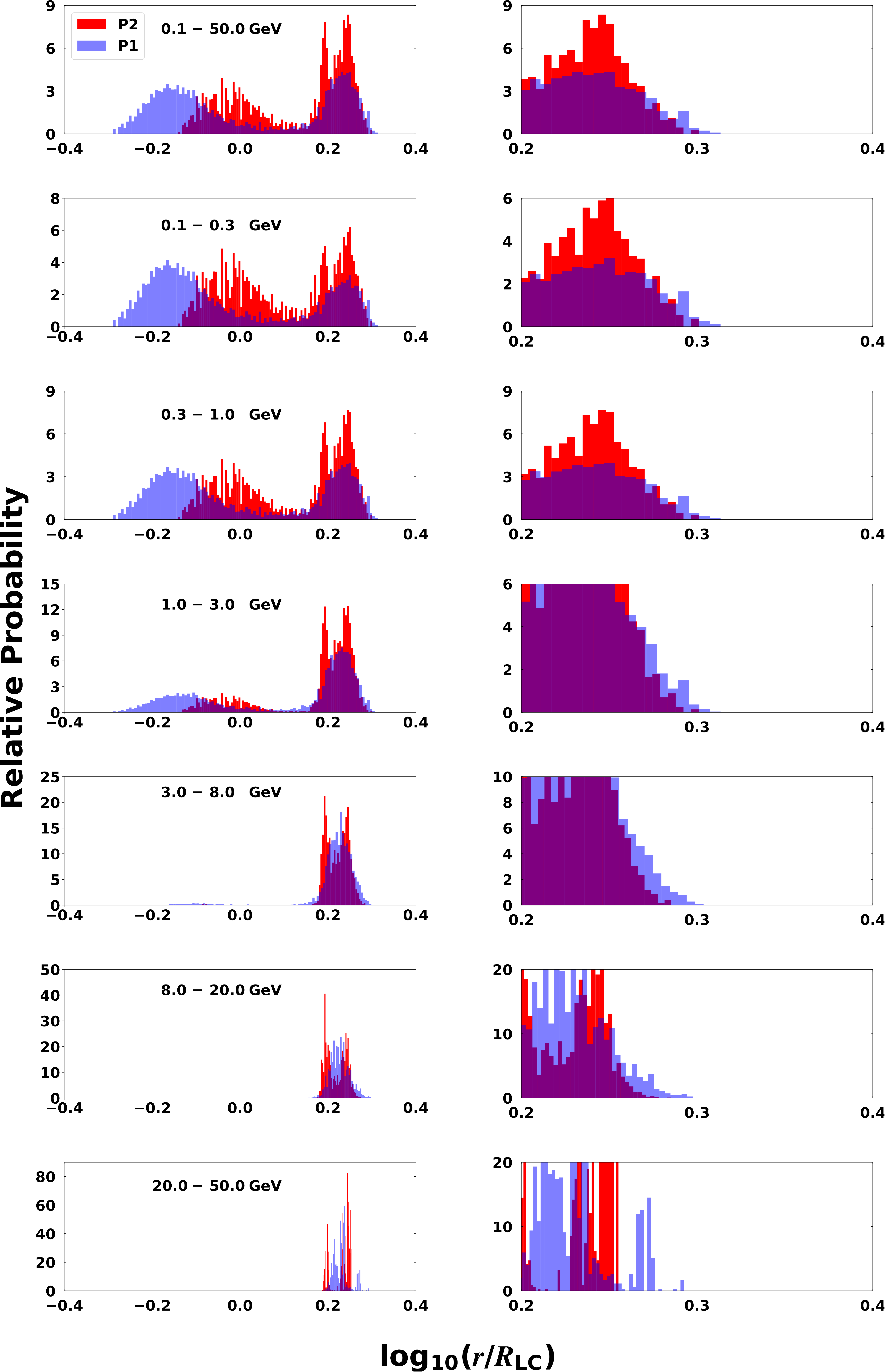}{0.45\textwidth}{(a)}
          \fig{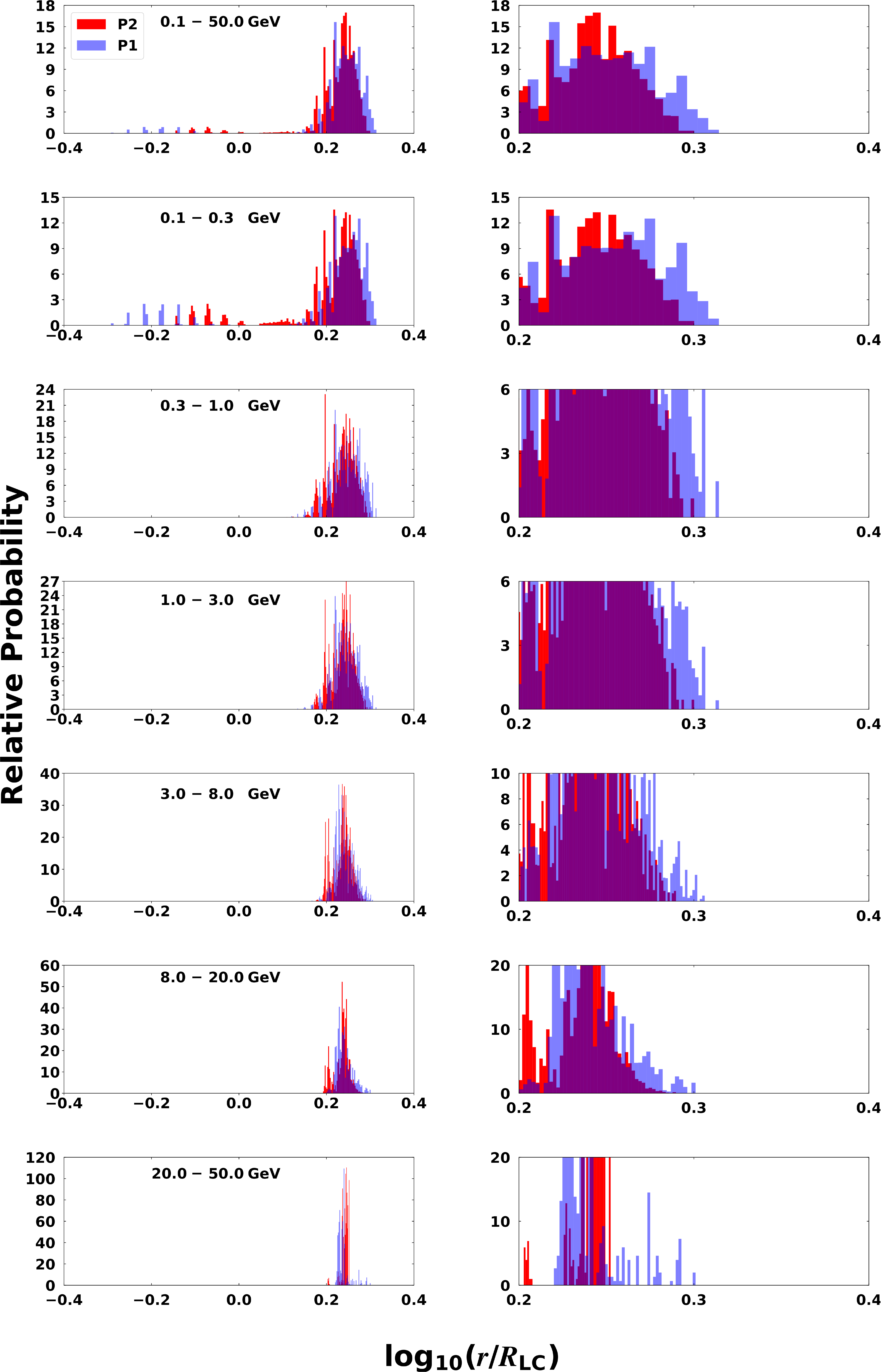}{0.45\textwidth}{(b)}
         }
\caption{The same as Figure~\ref{fig:hists_ecut} but for $\log_{10}(r/R_{\rm LC})$ (the emission radius). We note that the spatial grid of the FF $B$-field extends up to $r\approx 2R_{\rm LC}$, which explains the extent of the high-$r$ tail. \label{fig:hists_rem}}
\end{figure*}
In Figure~\ref{fig:hists_RlcLim1}a,~b, and~c, we limit the emission radius $r$ to altitudes at and beyond $R_{\rm LC}$ to investigate the change in the range of values for the pertinent quantities as compared to the previous cases where we considered emission from all altitudes. We show histograms for $\log_{10}$ of $E_{\gamma,\rm CR}$, $\rho_{\rm c}$, and $\gamma$, respectively, for the first scenario. At lower $E_\gamma$ (up to $\sim3$~GeV), the P1 bumps at lower values that we first noticed in Figure~\ref{fig:hists_ecut}a,~\ref{fig:hists_rho}a, and~\ref{fig:hists_gam}a are suppressed. This indicates that photons originating inside the light cylinder come from regions that are characterized by lower values of $E_{\gamma,\rm CR}$, $\rho_{\rm c}$, and $\gamma$.
This effect of limiting the emission altitudes is not as noticeable in the second scenario in Figure~\ref{fig:hists_RlcLim2}a,~b, and~c. 
For $E_{\gamma,\rm CR}$ the slight bump (including P1 and P2 emission) at low values of $E_{\gamma,\rm CR}$ disappears at lower $E_\gamma$ (up to $\sim{0.3}$~GeV).
For $\rho_{\rm c}$ and $\gamma$ the difference is insignificant, given the fact that the low-altitude $E$-field already suppresses the emission.

In Figure~\ref{fig:hists_rem} we show histograms for $\log_{10}(r/R_{\rm LC})$, for different energy ranges. We examined at what radial position the emission is coming from for each peak (in each scenario) in order to determine if the difference in $\rho_{\rm c}$ (between P1 and P2) is due to the different radial distances (as implied by the positive correlation between $\rho_{\rm c}$ and $r$ as seen in Figure~5 of \citealt{Kalapotharakos2014}) or just due to the different geometric properties of the particle trajectories. In the first scenario (left column), two bumps appear, for both peaks, at lower $E_\gamma$ (up to $\sim5$~GeV), situated around $\log_{10}(r/R_{\rm LC})\approx{-0.2}$ (for P1) ${-0.1}$ (for P2), and the second bump appears around ${0.2}$. The lower-altitude bump disappears with increasing $E_\gamma$ for both peaks. In (b), we show scenario~2, where there is a very small low-$E_\gamma$ bump (up to $\sim0.3$~GeV) at the same values of $\log_{10}(r/R_{\rm LC})$ as in scenario~1. The suppression of this bump is due to the low $E$-field assumed at low altitudes in scenario~2, suppressing particle acceleration and emission at these altitudes.
The $\log_{10}(r/R_{\rm LC})$ of P1 is very slightly larger than that of P2 for both scenarios, as seen in the zoom-ins, but not significantly so. Thus, we conclude that the systematically larger $\rho_{\rm c}$ for P2 cannot be solely ascribed to the fact that emission forming P2 originates at higher altitudes.

\section{Conclusions} \label{sec:concl}
There is an ongoing debate regarding the origin of the GeV emission detected from pulsars, with it being attributed either to CR or SR (or even IC; see \citealt{Lyutikov2012,Lyutikov2013}). One way in which to possibly discriminate between these options is to model the energy-dependent light curves and phase-resolved spectra of several bright pulsars.

We presented a refined calculation of the $\rho_{\rm c}$ of particle trajectories, impacting the CR loss rate and leading to smoother phase plots and light curves. However, this refinement had a rather small impact, as the broad structure of caustics and light curves remained similar to what was found previously. Additionally, we also found that the CRR limit was easily reached in the first scenario (constant $E$-field), and sometimes also in the second (two-step $E$-field).

We modeled $E_{\gamma}$-dependent light curves and spectra of the Vela pulsar in the HE regime assuming CR from primaries in an extended SG and current-sheet model to see if we can explain the origin of the decreasing P1/P2 ratio versus $E_\gamma$, expecting that the answer may lie in a combination of the values of geometric and physical parameters associated with each peak. Because the light curves probe geometry, e.g., $\alpha$, $\zeta$ and emission gap position and extent, while the spectrum probes both the energetics and geometry, we simultaneously fit both data sets with our model to obtain optimal fitting parameters.

We proceeded to isolate the P1/P2 effect by selecting photons that make up these two light curve peaks, and investigating the range of associated values of $E_{\gamma, \rm CR}$, $\rho_{\rm c}$, $\gamma$ and $r/R_{\rm LC}$. We found that the phase-resolved spectra associated with each peak indicated a slightly larger spectral cutoff for P2, confirming that P2 survives with an increase in energy, given its larger spectral cutoff. This was also seen in energy-dependent histograms of $E_{\gamma, \rm CR}$, confirming that this quantity was systematically larger for P2. The reason for this became clearer upon the discovery that both $\rho_{\rm c}$ and $\gamma$ were systematically larger for P2, for both scenarios. If CRR is reached, one expects $E_{\gamma, \rm CR}\propto \rho_{\rm c}^{1/2}$ for a constant $E$-field, so the larger $\rho_{\rm c}$ would explain the larger spectral cutoff for P2. Conversely, even if CRR is not attained, $E_{\gamma, \rm CR}\propto \gamma^3\rho_{\rm c}^{-1}$. Given the systematic dominance of $\gamma$ for P2 and the strong dependence of the third power, the larger spectral cutoff of P2 is thus explained by the larger $\gamma$. However, $\gamma$ is in principle larger on field lines where $\rho_{\rm c}$ is larger, because the CR loss rate $\propto \rho_{\rm c}^{-2}$ is smaller there, thus, the systematically larger values of $\rho_{\rm c}$. Moreover, the fact that the emission radii did not differ significantly for P1 and P2, underscores this conclusion: we are not observing a positive correlation between $\rho_{\rm c}$ and $r$ \citep[e.g.,][]{Kalapotharakos2014}, but at similar emission radii, $\rho_{\rm c}$ is systematically larger for P2. \textit{This means that the underlying $B$-field geometry}\footnote{It would be interesting to test this for other standard $B$-field geometries such as the retarded-vacuum field or an offset-dipole field, but such vacuum solutions preclude emission calculation beyond the light cylinder, muddling the issue.} \textit{at the emission location and for a given set of optimal pulsar parameters} is the fundamental reason why the spectral cutoff of P2 is larger than that of P1. In addition, we also found that the values of $\rho_{\rm c}$ and $\gamma$ remained larger for P2 when only considering emission beyond the light cylinder; in particular, the largest values of these quantities occurred there, pointing to dominant emission from that region to make up the GeV light curves.

We note that the drop in P1/P2 vs.\ $E_\gamma$ may not be universal, as also found by \citet{Brambilla2015}. We found a counterexample for a different choice of $\zeta_{\rm cut}$, where P1/P2 increases with $E_\gamma$. This was also the case for specific choices of the two-step acceleration $E$-field. There may also be other parameter combinations that can yield this behavior. However, this effect seems prevalent and has been seen in both HE and VHE data of bright pulsars.

In summary, we found reasonable fits to the energy-dependent light curves and phase-resolved spectra of Vela, and our model that assumes CR as the mechanism responsible for the GeV emission captures the general trends of the decrease of P1/P2 vs.\ $E_{\gamma}$, evolution/depression of the interpeak bridge emission, plus stable peak positions and a decrease in the peak widths as $E_{\gamma}$ is increased. However, an unknown azimuthal dependence of the $E$-field as well as uncertainty in the precise spatial origin of the emission, precludes simplistic discrimination of emission mechanisms. Similar future modeling of energy-dependent light curves and spectra within a striped-wind context that assumes SR to be the relevant GeV mechanism will be necessary to see if those models can also reproduce and explain these salient features in the case of Vela and other pulsars.

In the future, we can incorporate the SC radiation mechanism as done by \citet{Harding2018}. This mechanism seems to be able to produce spectra that are relatively higher at lower MeV energies (but this is not as relevant to the current paper) and that may provide better fits to the data. A future multiwavelength study of the evolution of P1/P2 with $E_\gamma$ could shed some more light on the underlying emission geometry and radiation mechanisms. For example, modeling of the VHE pulsed emission could scrutinize the general emission framework of any particular model, as well as constrain particle energetics.

\appendix
\section{Particle Transport and Refined calculation of the curvature radius $\rho_{\rm c}$} \label{sec:AppA}

In this appendix, we describe the updated procedure to calculate the radius of curvature of particle trajectories in the lab frame.

The motion of a charged particle in external $\boldsymbol{E}$ and $\boldsymbol{B}$-fields (in the absence of General Relativistic corrections) is described as follows
\begin{equation}
    \frac{d\boldsymbol{p}}{dt}=e\left[\boldsymbol{E}+\frac{1}{c}(\boldsymbol{v}\boldsymbol{\times} \boldsymbol{B})\right]+\boldsymbol{f}^{\rm rad},
\end{equation}
where $\boldsymbol{p}=\gamma m\boldsymbol{v}$ is the particle momentum and $\boldsymbol{f}^{\rm rad}$ is the radiative reaction force (e.g., \citealt{Landau1987}). 
Using the drift approximation that is correct to an accuracy of the order of the Larmor radius divided by positional radius $r$, the following equations may be
derived (e.g., \citealt{Sivukhin1965}):
\begin{eqnarray}
    \frac{dp_\parallel}{dt} & = & e\left(\boldsymbol{E}\cdot \boldsymbol{h}\right)+\frac{1}{2}p_\perp v_\perp {\rm div} \boldsymbol{h}+f^{\rm rad}_\parallel,\\
    \frac{d p_\perp}{dt} & = & -\frac{1}{2}p_\parallel v_\perp {\rm div} \boldsymbol{h}+f^{\rm rad}_\perp,
\end{eqnarray}
where $\boldsymbol{h}$ is a  unit vector parallel to the local $\boldsymbol{B}$-field, $p_\parallel$ and $p_\perp$ are the smoothed components of the momentum along and perpendicular to the local $\boldsymbol{B}$-field, $f_\parallel=(-2e^4/3m^2c^4)B^2\gamma^2\sin^2\psi\cos\psi$ and $f_\perp=f_\parallel\tan\psi$, with $\psi$ the pitch angle. In the limits of $\gamma\gg 1$ and $\psi\ll 1$, these two equations reduce to (see Eq.~[19] and [20] as well as Appendix~A of \citealt{Harding2005}, but dropping the cyclotron/synchrotron resonant absorption terms)
\begin{eqnarray}
  \frac{d\gamma}{dt} & = & \frac{eE_\parallel}{mc} -\frac{2e^4}{3m^3c^5}B^2p^2_\perp - \frac{2e^2\gamma^4}{3\rho_{\rm c}^2},\label{eq:CRloss}\\
  \frac{dp_\perp}{dt} & = & -\frac{3}{2}\frac{c}{r}p_\perp -\frac{2e^4}{3m^3c^5}B^2p^3_\perp\frac{1}{\gamma}.\label{eq:momperp} 
\end{eqnarray}
Thus, in these limits, Eq.~(\ref{eq:CRloss}) reduces to Eq.~(\ref{eq:transport1}), and we effectively drop the second equation, because we assume that the particle pitch angles remain close to zero. Therefore, while Eq.~(\ref{eq:momperp}) invokes the approximation ${\rm div}\boldsymbol{h}\simeq3/r$ that is valid close to the stellar surface (and this may need to be revised in future when assessing SR emission at large altitudes), it is not relevant to the current study, as we are only interested in the energetics of particles emitting CR.

While the first transport equation above determines the particle energetics, the trajectories of the high-energy emitting particles are assumed to always follow the asymptotic trajectories that are determined by the $\boldsymbol{E}\times\boldsymbol{B}$ drift as well as the $\boldsymbol{B}$-field structure (e.g., \citealp{Bai2010,Kalapotharakos2014,Gruzinov2012}). The particle velocity may thus be divided into a drift component and a component parallel to the local $\boldsymbol{B}$-field (see Eq.~[12] in \citealt{Kalapotharakos2014}):
\begin{equation} \label{eq:driftvelocity}
    \frac{\boldsymbol{v}}{c}=\frac{\boldsymbol{E}\times\boldsymbol{B}}{B^2+E_0^2}+f\frac{\boldsymbol{B}}{B},
\end{equation}
where $B^2_0-E_0^2=\boldsymbol{B}^2-\boldsymbol{E}^2$, $B_0E_0=\boldsymbol{E}\cdot\boldsymbol{B}$, and $E_0 \geq 0$. The factor $f$ is solved by assuming that the particle is moving close to the speed of light, formally setting $|\boldsymbol{v}|\rightarrow c$. Thus, while the direction of motion may be nearly tangent to the local $\boldsymbol{B}$-field at low altitudes, at $r \geq R_{\rm LC}$ the particle motion becomes predominantly radial (given the large drift component and the sweepback of the $\boldsymbol{B}$-field in the opposite, negatively toroidal, direction). This velocity may next be integrated to yield the particle trajectory (position).

To calculate the particle's trajectory as well as its associated $\rho_{\rm c}$, we first use a small, fixed step size $ds$ (arclength interval) along the particle trajectory in the lab frame. The particle is injected at the stellar surface, and we trace its motion. Because we are using a numerical solution of the FF $\boldsymbol{B}$-fields and $\boldsymbol{E}$-fields at a particular magnetic inclination angle $\alpha$ as in \citet{Harding2015} and \citet{Harding2018}, the three Cartesian components of the local $\boldsymbol{B}$-fields and $\boldsymbol{E}$-fields are available at any specified position (albeit they may have to be interpolated, given the chosen resolution of the numerical solution of the FF fields). One can thus use this information to map out a particle's trajectory according to Eq~(\ref{eq:driftvelocity}). Therefore, the particle positions ($x$,$y$,$z$) as well as the local first-order derivatives ($x^\prime$,$y^\prime$,$z^\prime$) along the trajectory as a function of the cumulative arclength $s$ (i.e., the normalized velocity) are used to compute both the full particle trajectory and its $\rho_{\rm c}(s)$.

The calculation involves three positions (previous, current, and next, denoted by indices $i-1$, $i$ and $i+1$, respectively). Let us denote the position at injection as $(x_{i-1},y_{i-1},z_{i-1})$.
The first-order derivatives $\boldsymbol{x}^\prime = d\boldsymbol{x}/ds = \boldsymbol{v}/c$ at this position is also available: $(x_{i-1}^\prime,y_{i-1}^\prime,z_{i-1}^\prime)$. We next step along the field line, updating the arclength $s$.
The position is then updated according to the Euler method:   
\begin{equation}
    x_i = x_{i-1} + x^\prime_{i-1}\cdot ds,
\end{equation}
and similarly for the other two spatial coordinates.
The current position and derivative $(x_{i}^\prime,y_{i}^\prime,z_{i}^\prime)$ are then saved. We similarly move to the next position,
\begin{equation}
    x_{i+1} = x_{i} + x^\prime_{i}\cdot ds,
\end{equation}
also for $y$ and $z$.
We thus have position and local direction components at three adjacent points with which we start the procedure. We step along the particle trajectory (this stepping procedure is repeated until some large radius is reached), so the $x_{i+1}$ becomes the current position and similar for the first-order derivative.

First, at the current position, we smooth the three spatial coordinates ($x$,$y$,$z$) using $s$ as the independent variable to counteract numerical noise or uncertainties that may be present in the numerical calculation of the global $B$-field structure (and also taking into account the spatial grid on which this $B$-field was calculated). The smoothing is performed using a KDE smoothing procedure involving a Gaussian kernel \citep{Parzen1962}. Consider the following parameters: $x_{\rm old}$, $x_{\rm new}$, $y_{\rm old}$, and $y_{\rm new}$, with the `old' and `new' values referring to the unsmoothed and smoothed variables, respectively. Because we smooth the positions, directions, and $\rho_{\rm c}$ as a function of $s$, $x_{\rm old}$ and $x_{\rm new}$ represents the arclength values. The smoothing procedure is implemented by the following set of equations, where we smooth using the full range of spatial steps along the particle trajectory
\begin{eqnarray}
A_j & = & \left(\frac{x_{{\rm old,}j}-x_{ {\rm new,}k}}{h}\right)^2\\
T & = & \sum_j{y_{{\rm old,}j}e^{-0.5A_j}}\\
B & = & \sum_j{e^{-0.5A_j}}\\
y_{{\rm new,}k} &=& T/B.
\end{eqnarray}

We choose the smoothing parameter $h$ as a fraction of $R_{\rm LC}$; this needs to be adapted when increasing or decreasing the step size. The smoothing parameter used in the KDE procedure sets the level of smoothing (i.e., the spatial range in $s$ over which smoothing occurs) and needs to be connected to $ds$ to avoid under- or oversmoothing. After some testing, we set $h=50 ds$.

Second, we noticed that the our use of a KDE smoothing procedure on the position coordinates introduced some artificial ``tails'' at low and high altitudes, thus, the procedure is failing at the edges of the position range. We thus piecewise match (using some small tolerance on the allowed fraction that the smoothed and unsmoothed positions may differ) the unsmoothed and smoothed spatial positions of the electron trajectory at particular $s$ values to get rid of these unwanted ``tails'' and to end up with the most satisfactory set of positions that constitute three smooth but realistic functions of arclength (i.e., a combination of the smoothed and unsmoothed positions as functions of $s$).

Third, we also smooth and then piecewise match the unsmoothed and smoothed directions of the electron trajectory at particular $s$ values to get rid of these unwanted ``tails'', as was done with the position coordinates.

Fourth, we use a second-order method involving interpolation by a Lagrange polynomial to obtain the second-order derivatives of the positions along the trajectory as a function of $s$, based on the (smoothed and matched) first-order derivatives \citep{Faires2002}:
\begin{equation}
x_i^{\prime\prime}(s) = \frac{\left(-3x^\prime_{i-1} + 4x^\prime_{i} - x^\prime_{i+1}\right)}{2ds},
\end{equation}
and similar for $y$ and $z$. We do this using both the smoothed and unsmoothed first-order derivatives of the position.

Fifth, with the second-order derivatives in hand, we calculate two instances of $\rho_{\rm c}$, one involving the unsmoothed (`us') and one involving the smoothed (`s') accelerations:
\begin{equation}
\rho_{\rm c,us}(s) = \frac{1}{\sqrt{x^{\prime\prime}_{\rm us}(s)^2+y^{\prime\prime}_{\rm us}(s)^2+z^{\prime\prime}_{\rm us}(s)^2}}, \\
\end{equation}
\begin{equation}
\rho_{\rm c,s}(s) = \frac{1}{\sqrt{x^{\prime\prime}_{\rm s}(s)^2+y^{\prime\prime}_{\rm s}(s)^2+z^{\prime\prime}_{\rm s}(s)^2}}. \\
\end{equation}
Finally, we piecewise match these two results for $\rho_{\rm c}(s)$ to get rid of ``tails'' in $\rho_{\rm c}$ at low and high altitudes, as before.

\section{Calculation of the phase-average and phase-resolved model spectra}\label{sec:AppB}

Formally, the phase-averaged spectrum is given by
\begin{equation}
\nu F_{\nu,\rm avg} = \frac{1}{d\Omega d^2}\int_0^{2\pi}\!\!\int_\zeta^{\zeta+d\zeta}E_{\gamma}^2 I_\gamma \sin\zeta d\zeta d\phi, \\
\end{equation}
with $I_\gamma =\dot{N_{\gamma}}/d\Omega dE_{\gamma}$, and $d\Omega = 2\pi\int_\zeta^{\zeta+d\zeta}\sin\zeta d\zeta = 2\pi\left[\cos\left(\zeta\right) - \cos\left(\zeta + d\zeta\right)\right].$ We note that the factor $d\Omega$ in front of the integral above does not cancel with this factor in $I_\gamma$, as the latter is calculated at each step along the particle trajectory, and then binned (so the solid angle information is effectively lost).

In our simulations, for a given phase bin (call it phase bin $j$), the instantaneous (phase-resolved) flux that would be measured if only considering the times corresponding to that bin for the purposes of instrument exposure is given by
\begin{equation}
\nu F_{\nu,j} \approx \frac{1}{d^2} E_{\gamma}^2 I_\gamma\left(\zeta+\frac{1}{2}d\zeta\right). \\
\end{equation}
Thus, the solid angle bins are considered small enough so that the $d\Omega$ in the prefactor cancels the $\sin\zeta d\zeta d\phi$ in the integrand. With this definition, the instantaneous flux (per bin) in a given phase bin can exceed the phase-averaged flux (i.e., $\nu F_{\nu,j}\gg \nu F_{\rm \nu,avg}$), particularly during the peaks of the light curve. To obtain the phase-averaged flux from our simulations, with a finite number of $\zeta$ and $\phi$ bins, we sum over the instantaneous fluxes and renormalize to the full phase interval
\begin{equation}
\nu F_{\nu,\rm avg} \approx \left(\frac{d\phi}{2\pi}\right)\sum_{j = 1}^{n_{\rm obs}} \nu F_{\nu,j} = \frac{1}{n_{\rm obs}}\sum_{j = 1}^{n_{\rm obs}} \nu F_{\nu,j},\\
\end{equation}
with $d\phi$ the constant observer phase bin width and $n_{\rm obs}$ the number of observer phase bins.

For phase-resolved analysis of a light curve feature, e.g., the light curve peaks with extent $\Delta\phi = \phi_{\rm high} - \phi_{\rm low}$ corresponding to multiple bins in our simulation, the flux is given by
\begin{equation}
\nu F_{\nu,\rm res} \approx \left(\frac{d\phi}{\Delta\phi}\right)\sum_{\phi_{\rm low}}^{\phi_{\rm high}}\nu F_{\nu,j}\approx \left(\frac{d\phi}{\Delta\phi}\right)\frac{1}{d^2} \sum_{j_1}^{j_2}E_{\gamma}^2 I_\gamma\left(\zeta+\frac{1}{2}d\zeta\right),
\end{equation}
with $j_1$ and $j_2$ the indices corresponding to $\phi_{\rm low}$ and $\phi_{\rm high}$.

\acknowledgments
This work is based on the research supported wholly / in part by the National Research Foundation of South Africa (NRF; Grant Numbers 87613, 90822, 92860, 93278, and 99072). The Grant holder acknowledges that opinions, findings and conclusions or recommendations expressed in any publication generated by the NRF supported research is that of the author(s), and that the NRF accepts no liability whatsoever in this regard. A.K.H.\ acknowledges the support from the NASA Astrophysics Theory Program. C.V.\ and A.K.H.\ acknowledge support from the \textit{Fermi} Guest Investigator Program.

\bibliography{VelaCRrho}{}
\bibliographystyle{aasjournal}

\end{document}